\documentclass[
preprint,
prd,
showpacs, 
amsmath, 
amssymb, 
floatfix,
nofootinbib,
]{revtex4}

\usepackage[dvips]{graphicx}

\begin{document}

\title{Apparent horizon formation in the head-on collision of gyratons}

\author{Hirotaka Yoshino}
\author{Andrei Zelnikov}
\author{Valeri P. Frolov}


\affiliation{Department of Physics, University of Alberta, 
Edmonton, Alberta, Canada T6G 2G7}

\preprint{Alberta-Thy-01-07}

\date{Submitted March 26, 2007; Published June 5, 2007}

%
%
\begin{abstract}

The gyraton model describes a gravitational field of an object moving
with the velocity of light which has finite energy and spin
distributed during some finite time interval $L$. 
A gyraton may be
considered as a classical toy model for a quantum wave packet of 
high-energy particles with spin.  In this paper we study a head-on
collision of two gyratons and black hole formation in this process.
The goal of this study is 
to understand the role of the gravitational
spin-spin interaction in the process of mini black hole formation in
particle collisions. To simplify the problem we consider several
gyraton models with special profiles of the energy and spin density
distribution. For these models we study the apparent horizon (AH)
formation on the future edge of a spacetime region before
interaction. We demonstrate that the AH forms
only if the energy duration
and the spin are smaller  than some critical values, while
the length of the spin distribution should be at least of the order of the
system gravitational radius.  We also study gravitational spin-spin
interaction in the head-on collision of two gyratons under the 
assumption that the values of gyraton spins are small. We demonstrate
that the metric in the interaction region for such gyratons depends
on the relative helicities of incoming gyratons,  and the collision of
gyratons with oppositely directed spins allows the AH formation in
a larger parameter region than in the collision of the gyratons with
the same direction of spins. 
Some applications of the obtained results to the mini
black hole production  at the Large Hadron Collider in TeV gravity
scenarios are briefly discussed.

\end{abstract}

\maketitle

%
%
\section{Introduction}

The black hole formation in high-energy particle collisions is an
important issue especially in the context of TeV gravity
scenarios~\cite{ADD98, RS99-1, RS99-2}.  In the theory with large
extra dimensions, the Planck energy could be of the order of TeV and
collisions of particles with the center-of-mass energy greater
than the Planckian one will occur in future accelerators such as the
Large Hadron Collider (LHC) at CERN \cite{BHUA}. A detailed study of
mini-black-hole production, especially at the threshold of this effect,
would require the complete theory of quantum gravity. However, if
the mass of a created black hole is much larger than the Planck mass, one
can use a semiclassical approximation to describe such processes. In
this approximation, the black hole formation in particle collision 
and its subsequent decay in the process of the Hawking evaporation 
are studied in the framework of the (semi)classical general relativity.  
The apparent horizon (AH) is a useful
tool for estimation of black hole production rate in this
approximation, because the existence of an AH is a sufficient
condition for the black hole formation.

The first work along this line was done by Eardley and Giddings
\cite{EG02} in the four-dimensional case.  Since the high-energy
particles are relativistic, they used the Aichelburg-Sexl (AS) metric 
\cite{AS71} to describe the gravitational field of such particles
before their collision. The AS metric can be obtained  by boosting a
Schwarzschild black hole to the speed of light  and keeping the
energy $p$ of the boosted black hole fixed.  The gravitational field
of the AS particle is a shock and it is localized on the null plane
($u=0$ for one of the particles and $v=0$ for the other one). One of
the null generators on each of the null planes represents a particle
trajectory, while its gravitational field is distributed in the
transverse plane orthogonal to the direction of motion. Two AS
particles do not interact before the collision and the metric outside
of the interacting region is known explicitly. Eardley and Giddings
analytically studied the AH on some slice  ($u=0\ge v$ and $v=0\ge
u$) and derived the maximal impact parameter $b_{\mathrm{max}}$  for
the AH formation.  The quantity $\sigma_{\rm AH}=\pi
b_{\mathrm{max}}^2$ gives the lower bound on the cross section
of the black hole production.

The results of \cite{EG02} were generalized  by one of us and Nambu
\cite{YN03} for the mini-black-hole formation in the
higher-dimensional spacetimes. In this work  the numerical
calculations were used. Later, one of us and Rychkov \cite{YR05}
improved these results by studying the AH on the futuremost
slice that can be adopted without the knowledge of the interacting
region (i.e., $u=0\le v$ and $v=0\le u$). Using this approach the
stronger lower bound  $\sigma_{\rm AH}/\pi[r_h(2p)]^2 \simeq 2$--$3$
on the cross section  of the black hole production in the collision of
AS particles was obtained for spacetime dimensionality $D=5$--$11$.

Certainly the model of colliding AS particles is oversimplified.  The
AS particles are assumed to be neutral and spinless. In reality all
the known elementary particles have spin and most of them have either
electric or color charge as well. 

Charged particles have additional charge-charge interaction
besides the gravitational interaction. Moreover their gravitational
interaction can also be modified. The latter effect was  discussed to
some extent by one of us and Mann \cite{YM06}. In that paper a
boosted Reissner-Nordstr\"om metric was used as the model of an
ultrarelativistic charged particle and the head-on collision in a spacetime
with an arbitrary number of dimensions was studied. The results
obtained in that paper indicate that the charge makes the AH formation
more difficult. It was also argued  that the effects of
the quantum electrodynamics could change the results. 
The results of \cite{YM06}
were later used by Gingrich \cite{G06} who reconsidered the black
hole production rate at the LHC.

In the quantum mechanical description, the colliding particles are
characterized by wave packets which have finite duration in time
\cite{GR04}.  To take into account this effect as well as to include
spin-spin interaction,  in this paper we study head-on collisions
of two gyratons.

The gyraton model was proposed in \cite{FF05}. The motivation of this
paper was to find the gravitational  field generated 
by a beam pulse of spinning radiation with a finite time duration,
which is propagating at the speed of light. In the
gyraton model, the metric outside of the source satisfies the vacuum
Einstein equations and the gravitational field is distributed in the
plane transverse to the direction of motion. Unlike an AS particle,
the gravitational field of a gyraton is not a shock wave but has the
finite duration in time. A gyraton may have spin which manifests
itself in the dragging-into-rotation effect.
The AS particle metrics can be obtained from the gyraton solutions if
the duration is taken infinitely small and the spin vanishes.
General properties of gyraton metrics were studied  in detail in
\cite{FIZ05}. Electrically charged gyratons and gyratons in the
supergravity were discussed in \cite{FZ06, FF06}, respectively.
Gyraton solutions can be also generalized to the case when the
spacetime is asymptotically anti-de Sitter \cite{FZ05,CKZ06}. 

Colliding gyratons which we consider in this paper differ from the
AS particles both by the presence of spin and the finite duration in
time. Let us discuss briefly what kind of new features one can
expect. First natural question is: Can one include spin effects in
the interaction between highly nonrelativistic particles by  boosting the
Kerr metric? Such boosted Kerr black hole solutions were considered,
e.g. in \cite{LS92,BN,BH03}, in the four-dimensional spacetime and in
\cite{Y04} in higher dimensions. The main problem in this approach is
the following. In order to have a well-defined limit for the boosted
metric, one needs to keep the energy $p$ of the system fixed, so that
the mass of the black hole $M=\gamma^{-1}p$ must vanish when the
$\gamma$-factor infinitely grows. If one assumes that the spin $s$
remains finite in this limit, the rotation parameter $a=s/M$
infinitely grows, so that the metric describes a naked singularity.
The radius of the ring singularity is of the order of
$a$ and also infinitely grows.  The latter problem can be avoided
by assuming that the rotation parameter $a$ remains finite in the
infinite boost limit, as it was done in the above references.
Although finite results different from the AS particle 
can be obtained by this procedure,
fixing $a$ means that the spin $s=aM$ of the boosted object vanishes. 
Furthermore, in this limit we have an object of typical size $a$, 
which does not satisfy the requirement that we
would like to have a point-like object. 
Thus a boosted  Kerr black hole does not provide one with  a suitable model
for an ultrarelativistic particle with spin, e.g. for a photon. In the
gyraton model the spin is easily included. 

There is another aspect of the high-energy particle collision which
the gyraton model may help to understand better. Recently the
validity of an AS particle as the model of a high-energy elementary
particle was questioned in \cite{R04}. In this model the curvature
invariants at the moment of the collision of the two planes,
representing the colliding particles, are infinite, so that formally
higher-curvature quantum corrections may 
be important. This
problem was discussed in  \cite{GR04}. It was argued that the
quantum effects, such as the finite size and finite duration in time
of the incoming wave packets, can help to solve this problem. The
quantum-to-classical transition in the description of the mini-black-hole 
formation in the particle collision is an interesting open
question. We do not address it in our paper, but instead we use the
gyraton model in order to discuss how the finiteness of the duration in
time of the colliding objects modifies the results of the AS
approximation. In such an approach, the gyratons might be considered as
an effective model for the quantum wave packets.  

With these motivations we study the AH formation
in the head-on collision of gyratons. For simplicity we
consider the four-dimensional case. The paper is organized as follows.
In the next section, 
we introduce several gyraton models:
a gyraton without spin, its AS limit, and spinning gyratons.
As for spinning gyratons, we consider two 
types depending on relative locations
of the energy and spin profiles.
Then we set up five cases of head-on collisions of gyratons.
In Sec. III, we derive an AH equation
on the future edge of the spacetime region before 
interaction, $u=0\le v$ and $v=0\le u$. 
We present the numerical results
of this study in Sec. IV. The conditions on the spin and on
the energy and spin durations for the AH formation 
are obtained for each of the collision cases.
Then in Sec. V we focus our attention on the study of
the spin-spin interaction.
In general, this is a complicated problem, since it requires the
knowledge of the metric in the region of interaction. To obtain it, 
one needs to solve nonlinear Einstein equations. We simplify
the problem by assuming that the spins of the interacting objects are
small and solve the equations by using a method of perturbation.
Then we study again the AH formation on the new slice
that is the future edge of the solved region. In the adopted
approximation we obtain spin-spin interaction corrections to 
the mini-black-hole production.
Sec. VI contains a summary of the results and a discussion
of their possible applications for the study of mini-black-hole production
at the LHC.

%
%
\section{System setup}

In this section, we set up the problem of the 
head-on collisions of two gyratons.
We first review the gyraton model
in a four-dimensional spacetime in Sec. II A. 
The gyraton has the total energy $p$ and the spin $J$
which are distributed in time. Their time profiles  
are characterized by two functions. 
We introduce four kinds of gyratons by specifying these two functions.
Next in Sec. II B, we introduce a system of null geodesic coordinates,
which is necessary for specifying the slice to study the AH existence.
It is also useful for clarifying the gravitational field of gyratons.
In Sec. II C, we set up five cases of head-on collisions of two gyratons,
using the introduced four gyratons.

%
%
\subsection{The gyraton model}

The gyraton model was proposed in \cite{FF05}. In that paper,
the gravitational field generated by a beam pulse of spinning radiation
was first studied in the weak field approximation
and then the exact solutions of the Einstein equations
were obtained, which reduce to the
approximate solution at far distance from the source.
These solutions were obtained in any number of spacetime dimensions. 
In particular, a four-dimensional gyraton has the metric
%
\begin{equation}
ds^2=-d\bar{u}d\bar{v}+d\bar{r}^2+\bar{r}^2 d\bar{\phi}^2
-4G\left(2p\log\bar{r}\chi_p(\bar{u})d\bar{u}-J\chi_j(\bar{u})d\bar{\phi}\right)d\bar{u}.
\label{gyraton-metric-original}
\end{equation}
%
This metric represents a spacetime in which a segment-shaped source
located at $\bar{r}=0$
with the total energy $p$ and the spin $J$ is propagating
at the speed of light along $\bar{u}=\mathrm{const}$.
The existence of the term $d\bar{\phi}d\bar{u}$ in 
Eq.~\eqref{gyraton-metric-original} 
indicates presence of a dragging-into-rotation effect
generated by the spin source.
The functions $\chi_p(\bar{u})$ and $\chi_j(\bar{u})$
are arbitrary functions satisfying the normalization conditions
%
\begin{equation}
\int\chi_p(\bar{u})d\bar{u}=\int\chi_j(\bar{u})d\bar{u}=1.
\end{equation}
%
They represent the energy and spin density profiles,
respectively.

Introducing the new coordinates
$\bar{\boldsymbol{x}}:=(\bar{r}\cos\bar{\phi},\bar{r}\sin\bar{\phi})$, 
non-zero components of 
the energy-momentum tensor of the gyraton are calculated as
%
\begin{equation}
T_{\bar{u}\bar{u}}=p\chi_p(\bar{u})\delta(\bar{\boldsymbol{x}})
+\pi GJ^2\chi_j^2(\bar{u})\delta^2(\bar{\boldsymbol{x}}),
\label{Tuu}
\end{equation}
%
%
\begin{equation}
T_{\bar{u}{a}}=\frac{J}{4}\chi_{j}(\bar{u})
\epsilon_{{a}{b}}\partial_{{b}}\delta(\bar{\boldsymbol{x}}).
\label{Tua}
\end{equation}
%
These formulas show that the source has an infinitely narrow shape.
For a realistic beam pulse, the source should have a finite radius 
$\bar{r}=\bar{r}_{\rm s}$ and the metric will
take a different form from Eq.~\eqref{gyraton-metric-original}
for $\bar{r}<\bar{r}_{\rm s}$.
Therefore Eq.~\eqref{gyraton-metric-original}
is considered to be the metric outside of the
source $\bar{r}\ge \bar{r}_{\rm s}$.
In this paper, we do not take account of this finiteness (in space) 
of the beam pulse for simplicity and
adopt Eq.~\eqref{gyraton-metric-original} for arbitrary values of $\bar{r}$
(i.e., $\bar{r}_{\rm s}=0$).

Hereafter we adopt the gravitational radius of
$2p$, i.e. $r_h(2p)=4Gp$, as the unit of the length.
In this length unit, the gyraton metric is represented as
%
\begin{equation}
ds^2=-d\bar{u}d\bar{v}+d\bar{r}^2+\bar{r}^2 d\bar{\phi}^2
-2\log\bar{r}\chi_p(\bar{u})d\bar{u}^2+2j\chi_j(\bar{u})d\bar{\phi}d\bar{u}.
\end{equation}
%
Here, $j$ is a dimensionless quantity defined by  $j:=J/2pr_h(2p)$
and we use $j$ as a parameter to specify the spin of the gyraton.

\begin{figure}[tb]
\centering
{
\includegraphics[height=0.2\textheight]{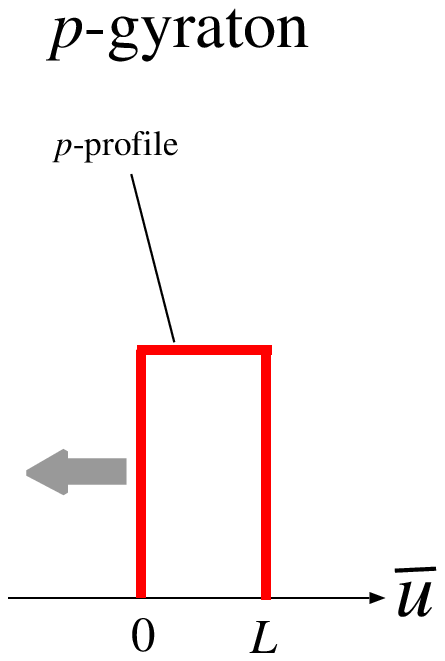}
\hspace{5mm}
\includegraphics[height=0.2\textheight]{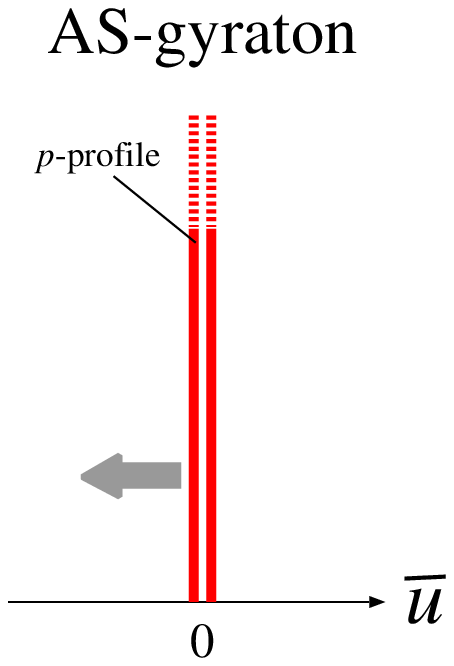}
\hspace{5mm}
\includegraphics[height=0.2\textheight]{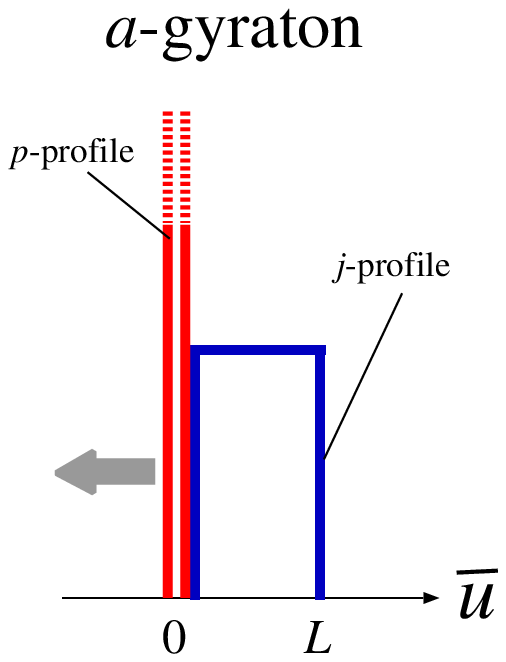}
\hspace{5mm}
\includegraphics[height=0.2\textheight]{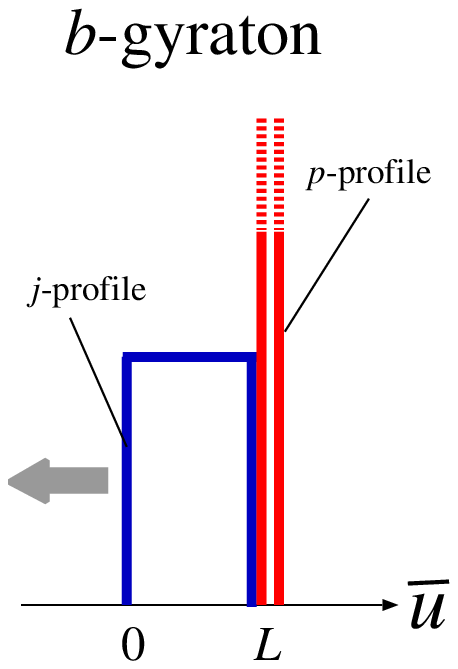}
}
\caption{The energy and spin density profiles, $\chi_p(\bar{u})$ and $\chi_j(\bar{u})$,
for the {\it p}--, AS--, {\it a}--, and {\it b}-- gyratons, respectively. The gray arrows
show the directions of propagation.}
\label{p-j-profiles}
\end{figure}

The gyraton model is specified by determining the functions
$\chi_p(\bar{u})$ and $\chi_j(\bar{u})$.
The interaction between two gyratons with arbitrary profiles 
$\chi_p(\bar{u})$ and $\chi_j(\bar{u})$
is a quite complicated problem which, in the general case,
requires numerical calculations.
Hence it is natural to consider first the simplest profiles 
for which the null geodesic coordinates can be studied analytically. 
For this reason, in this paper we consider four types of gyratons
whose energy and spin profiles
are as shown in Fig.~\ref{p-j-profiles}.
We will explain them one by one in the following.
For convenience, we introduce the following function
%
\begin{equation}
\vartheta(\bar{u},L)=
\frac{1}{L}
\left(\theta(\bar{u})-\theta(\bar{u}-L)\right)\ ,
\end{equation}
%
where $\theta(\bar{u})$ is the Heaviside step function.
Its integral over $\bar{u}$ is 1, and in the limit $L\to 0$ it gives a
$\delta$-function.

%
%

%
%
\subsubsection{$p$--gyraton}

The first one is a gyraton without spin with energy duration $L$. 
For this model, we adopt
%
\begin{equation}
\begin{cases}
\chi_p(\bar{u})=\vartheta(\bar{u},L),
\\
\chi_j(\bar{u})=0.
\end{cases}
\label{chip-chij-withoutspin}
\end{equation}
%
This model is useful for studying the effect of the energy duration
on the AH formation. We simply call it a ``spinless gyraton'' 
or a ``$p$--gyraton'' because it has only one parameter, energy $p$.

%
%
\subsubsection{AS--gyraton}

The second one is an Aichelburg-Sexl (AS) particle \cite{AS71}
with
%
\begin{equation}
\begin{cases}
\chi_p(\bar{u})=\delta(\bar{u}),\\
\chi_j(\bar{u})=0.
\label{chip-chij-AS}
\end{cases}
\end{equation}
%
This is the limit $L\to 0$ of the $p$-gyraton.
Hereafter we call it an ``AS--gyraton'' for short.

%
%
\subsubsection{{\it a}--gyraton and {\it b}--gyraton}

The remaining two gyratons, which are referred to as 
an ``{\it a}--gyraton'' and a ``{\it b}--gyraton'', 
have nonzero spin.
We adopt the following functions of $\chi_p(\bar{u})$ and $\chi_j(\bar{u})$
for {\it a}-- and {\it b}-- gyratons:
%
\begin{eqnarray}
&\textrm{{\it a}--gyraton:} & 
\begin{cases}
\chi_p(\bar{u})=\delta(\bar{u}),\\
\chi_j(\bar{u})=\vartheta(\bar{u},L); 
\end{cases}
\label{chip-chij-typea}
\\ 
&\textrm{{\it b}--gyraton:} & 
\begin{cases}
\chi_p(\bar{u})=\delta(\bar{u}-L),\\
\chi_j(\bar{u})=\vartheta(\bar{u},L),
\end{cases} 
\label{chip-chij-typeb}
\end{eqnarray}
%
In these two models, $L$ represents the spin duration and
the energy has zero duration. 
We call them an {\it a}--gyraton and a {\it b}--gyraton, respectively,
because for the {\it a}--gyraton, 
the spin source comes {\it after} the energy source,
while for the {\it b}--gyraton,
the spin source comes {\it before} the energy source. 
These two models are useful for studying the effect of the spin and its 
density duration on the AH formation. By comparing the results of 
{\it a}-- and {\it b}-- gyratons,
we can understand to what extent the obtained results depend on 
relative positions of the energy and spin
density profiles. Each of the gyratons is reduced to an AS--gyraton
if we take $j=0$.

Readers might wonder why we do not adopt 
$\chi_p(\bar{u})=\chi_j(\bar{u})=\vartheta(\bar{u},L)$ for spinning gyratons. 
This is because of a technical problem. In the next subsection,
we derive a coordinate system 
based on the null geodesic congruences.
This coordinate system can be analytically derived 
for {\it a}-- and {\it b}-- gyratons,
but not for $\chi_p(\bar{u})=\chi_j(\bar{u})=\vartheta(\bar{u},L)$.

%
%
\subsection{Null geodesic coordinates}

In this subsection, we introduce null geodesic coordinates, 
which are very useful for specifying the slice on which we study the AH.
We introduce new coordinates $(u,v,r,\phi)$ by
%
\begin{equation}
\begin{array}{ccl}
\bar{u}&=&u,\\
\bar{v}&=&v+F(u,r),\\
\bar{r}&=&G(u,r),\\
\bar{\phi}&=&\phi+H(u,r).
\end{array}
\end{equation}
%
We assume that the two coordinate systems coincide
for $\bar{u}=u\le 0$ and hence $F=H=0$ and $G=r$ for $u\le 0$.
For $u\ge 0$, we require a line $v, r, \phi=\mathrm{const.}$ 
to be a null geodesic
and the coordinate $u$ to be its affine parameter.
This requirement is realized if and only if the 
following relations are satisfied:
%
\begin{equation}
H_{,u}=-\frac{j\chi_j(u)}{G^2},
\label{eq-Hu}
\end{equation}
%
%
\begin{equation}
F_{,u}=G_{,u}^2-2\chi_p(u)\log G-\frac{j^2\chi_j^2(u)}{G^2},
\label{eq-Fu}
\end{equation}
%
%
\begin{equation}
F_{,r}=2G_{,u}G_{,r}.
\label{eq-Fr}
\end{equation}
%
These relations determine $F$, $G$, and $H$.
In terms of these functions 
the metric takes the form
%
\begin{equation}
ds^2=-dudv+G_{,r}^2dr^2+G^2(d\phi+H_{,r}dr)^2.
\label{metric-nullgeodesic}
\end{equation}
%
Eliminating $F$ from Eqs. \eqref{eq-Fu} and \eqref{eq-Fr},
we find
%
\begin{equation}
G_{,uu}=-\frac{\chi_p(u)}{G}+\frac{j^2\chi_j^2(u)}{G^2}.
\label{equation-for-G}
\end{equation}
%
Once this equation is solved, one can find $H$ by solving
Eq.~\eqref{eq-Hu} and 
determine all the coefficients in
the metric \eqref{metric-nullgeodesic}.

%
%
\subsubsection{p--gyraton and AS--gyraton}

For a spinless gyraton,
using Eq.~\eqref{chip-chij-withoutspin}, we find
%
\begin{equation}
G(u,r)=
\begin{cases}
\tilde{G}(u,r),
&
(0\le u\le L),
\\
\tilde{G}_{,u}(L,r)(u-L)+\tilde{G}(L,r),
&
(L\le u),
\end{cases}
\label{G-formula-withoutspin}
\end{equation}
%
where
%
\begin{equation}
\tilde{G}(u,r):=
r\exp
\left(
-\left[\mathrm{erf}^{-1}
\left(y\right)
\right]^2
\right),
\label{tildeG-formula-withoutspin}
\end{equation}
%
%
\begin{equation}
y:=\sqrt{\frac{2}{\pi L}}\frac{u}{r}.
\label{def-y}
\end{equation}
%
Here, the function $\mathrm{erf}^{-1}(y)$ denotes the inverse function
of the error function: $x=\mathrm{erf}^{-1}(y)$ is equivalent to
%
\begin{equation}
y=\mathrm{erf}(x):=\int_0^x\frac{2}{\sqrt{\pi}}\exp(-t^2)dt.
\end{equation}
%
In the limit $L\to 0$, the function $G$ reduces to
%
\begin{equation}
G(u,r)=r-u/r,~~(0\le u),
\label{G-formula-AS}
\end{equation}
%
and the metric \eqref{metric-nullgeodesic} coincides with the AS-gyraton
in the null geodesic coordinates \cite{R04,YR05}.

%
\begin{figure}[tb]
\centering
{
\includegraphics[width=0.35\textwidth]{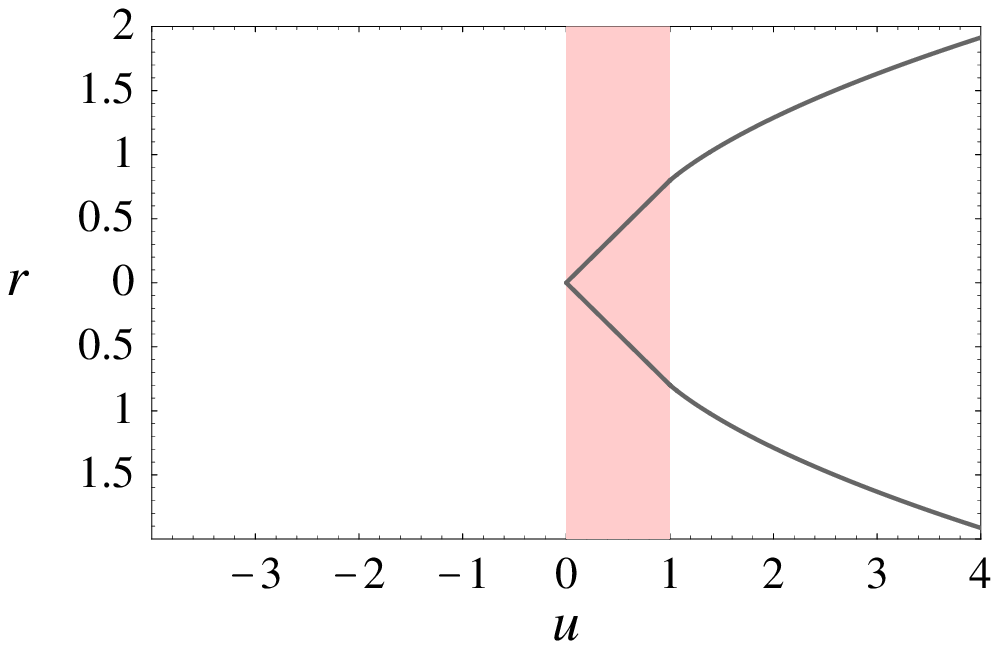}
\hspace{10mm}
\includegraphics[width=0.35\textwidth]{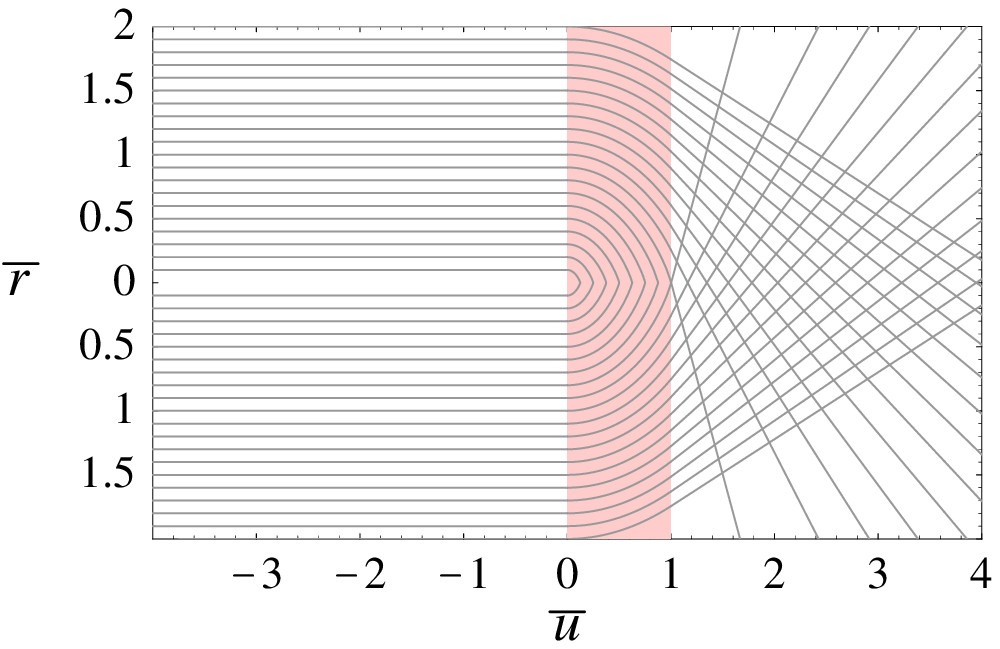}
}
\caption{The left plot shows the coordinate singularity
in the $(u,r)$ coordinates where
$G=0$ for a $p$--gyraton with $L=1$.
The lines $r=\mathrm{const.}$ are null geodesics and 
they hit the coordinate singularity.
The right plot shows the light rays in the $(\bar{u}, \bar{r})$ coordinates.
The light rays with an identical $r$ value bend due to an
attractive force and focus to one point on a symmetry axis,
which corresponds to $G=0$.  }
\label{nullray-nospin}
\end{figure}
%

We should point out that there is a coordinate singularity at
%
\begin{equation}
u=
\begin{cases}
\sqrt{{\pi L}/{2}}\ r, 
&
(0\le u\le L),
\\
L-\tilde{G}(L,r)/\tilde{G}_{,u}(L,r),
&
(L\le u),
\end{cases}
\end{equation}
%
where $G=0$. The shape of the singularity is 
shown by a solid line at the left plot of Fig.~\ref{nullray-nospin}.
In order to understand the meaning of this singularity, it is useful to consider
null geodesics $v,r,\phi=\mathrm{const.}$ These null geodesics plunge into
the coordinate singularity. Let us go back to the
original $(\bar{u}, \bar{v}, \bar{r}, \bar{\phi})$
coordinates. The trajectories of the light rays
in the coordinates $(\bar{u}, \bar{r})$ are shown in the right plot of 
Fig.~\ref{nullray-nospin}. Because of the gravitational effect of the gyraton energy,
the proper circumference of a congruence 
of light rays with an identical $r$ value 
becomes small as $\bar{v}$ is increased and 
eventually becomes zero. This is 
where the congruence hits the coordinate singularity in the $(u,v,r,\phi)$
coordinates. Thus, the coordinate singularity corresponds to the symmetry axis
and therefore we call it the {\it focusing singularity}.

%
%
\subsubsection{{\it a}--gyraton}

%
\begin{figure}[tb]
\centering
{
\includegraphics[width=0.35\textwidth]{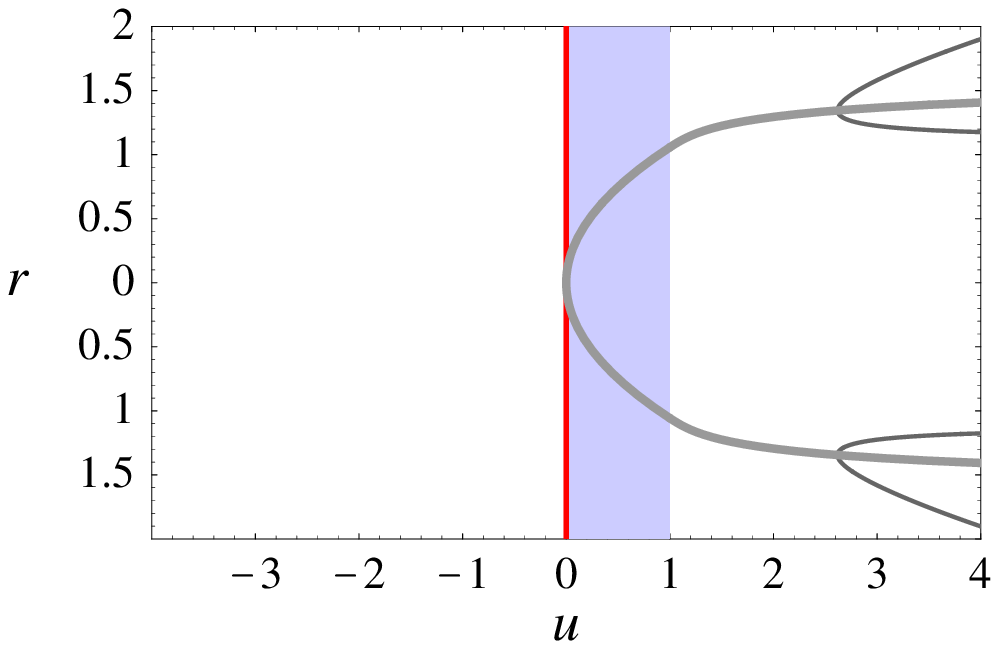}
\hspace{10mm}
\includegraphics[width=0.35\textwidth]{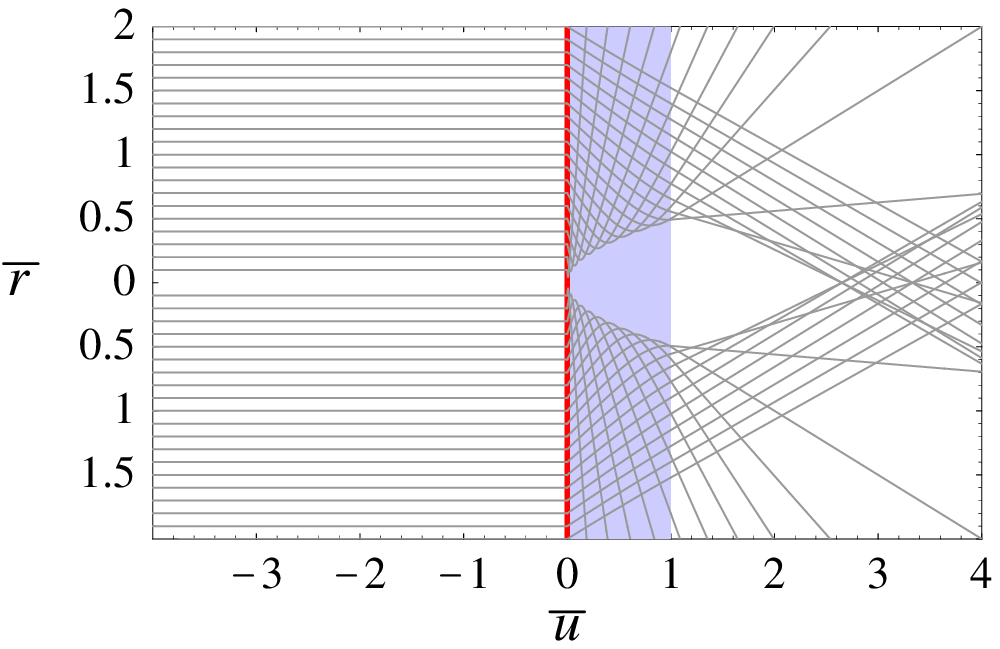}
}
\caption{The left plot shows the coordinate singularities
in the $(u,r)$ coordinates where 
$G=0$ (dark gray lines) and $G_{,r}=0$ (light gray line)
for an {\it a}--gyraton with $L=1$.
The lines $r=\mathrm{const.}$ are null geodesics and they hit one of 
the two coordinate singularities. The right plot shows the
propagation of light rays in the $(\bar{u}, \bar{r})$ coordinates.
Light rays with a large $r$ value focus to one point, 
which corresponds to $G=0$.
On the other hand, light rays with a small $r$ value
bend outward due to a repulsive force around the center. Then
two neighboring light rays cross each other, which corresponds to $G_{,r}=0$. }
\label{nullray-advance}
\end{figure}
%

Now we turn to the spinning {\it a}--gyraton.
Using Eq.~\eqref{chip-chij-typea}, Eqs.~\eqref{equation-for-G}
and \eqref{eq-Hu}
are solved as
%
\begin{equation}
G(u,r)=\begin{cases}
\tilde{G}(u,r),
&
(0\le u\le L),
\\
\tilde{G}_{,u}(L,r)(u-L)+\tilde{G}(L,r),
&
(L\le u);
\end{cases}
\label{G-formula-typea}
\end{equation}
%
%
\begin{equation}
H(u,r)=\begin{cases}
\tilde{H}(u,r),
&
(0\le u\le L),
\\
\tilde{H}(L,r),
&
(L\le u),
\end{cases}
\label{H-formula-typea}
\end{equation}
%
%
where
\begin{equation}
\tilde{G}(u,r):=r\sqrt{\frac{j^2/L^2+\left[(1+j^2/L^2)(u/r^2)-1\right]^2}{1+j^2/L^2}};
\label{tildeG-formula-typea}
\end{equation}
%
%
\begin{equation}
\tilde{H}(u,r):=\arctan \frac{j/L}{1-r^2/u}.
\label{tildeH-formula-typea}
\end{equation}
%
In this case, there are two coordinate singularities.
One is the singularity at
%
\begin{equation}
u=\frac{r^2(1-r^2/L)}{1+j^2/L^2-r^2/L}, ~~(L\le u),
\end{equation}
%
where $G=0$, and the other is at
%
\begin{equation}
u=\begin{cases}
\left(1+j^2/L^2\right)^{-1/2} r^2,
&
(0\le u\le L),
\\
L-\tilde{G}_{,r}(L,r)/\tilde{G}_{,ur}(L,r),
&
(L\le u),
\end{cases}
\end{equation}
%
where $G_{,r}=0$. 
The two coordinate singularities are shown in 
the left plot of Fig.~\ref{nullray-advance}.
A light ray $v,r,\phi=\mathrm{const.}$ plunges into one of the two singularities.
The propagation of light rays 
in the $(\bar{u}, \bar{r})$ coordinates
is shown in the right plot
of Fig.~\ref{nullray-advance}.
Because there is the energy distribution at $\bar{u}=0$, 
the light rays bend quickly there. Then the light rays with a sufficiently large 
$r$ value focus to one point
and this is the focusing singularity $G=0$. On the other hand, 
around the center, the gravitational field generated by the spin source is repulsive
and the light rays of a small $r$ value bend outward. 
Because of this effect, the neighboring 
light rays cross with each other and this is where the congruence 
hits the coordinate singularity $G_{,r}=0$.
Therefore we call it the {\it crossing singularity}.

%
%
\subsubsection{{\it b}--gyraton}

%
\begin{figure}[tb]
\centering
{
\includegraphics[width=0.35\textwidth]{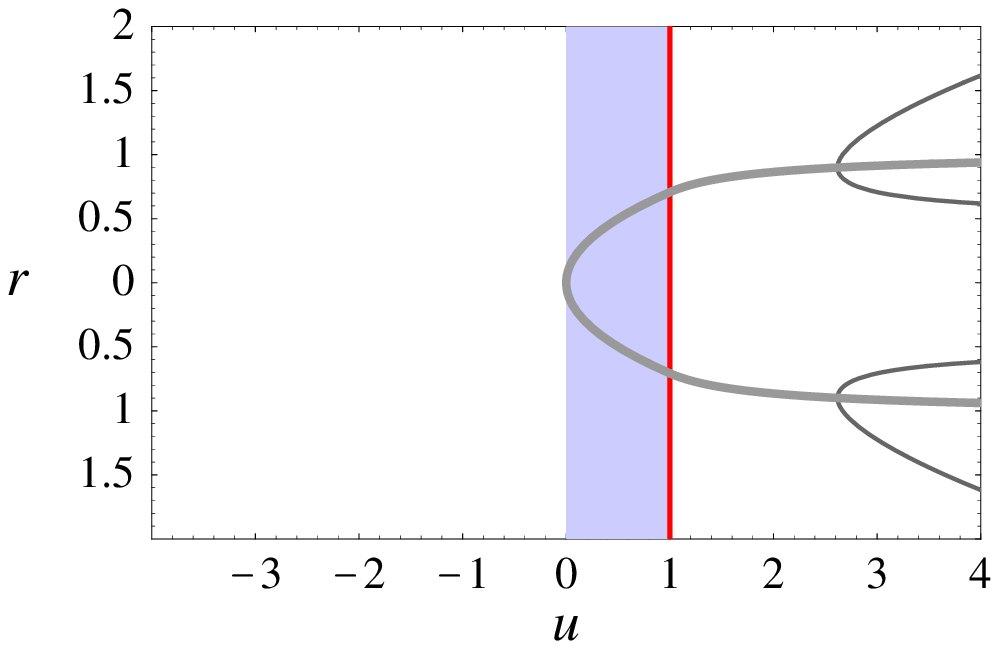}
\hspace{10mm}
\includegraphics[width=0.35\textwidth]{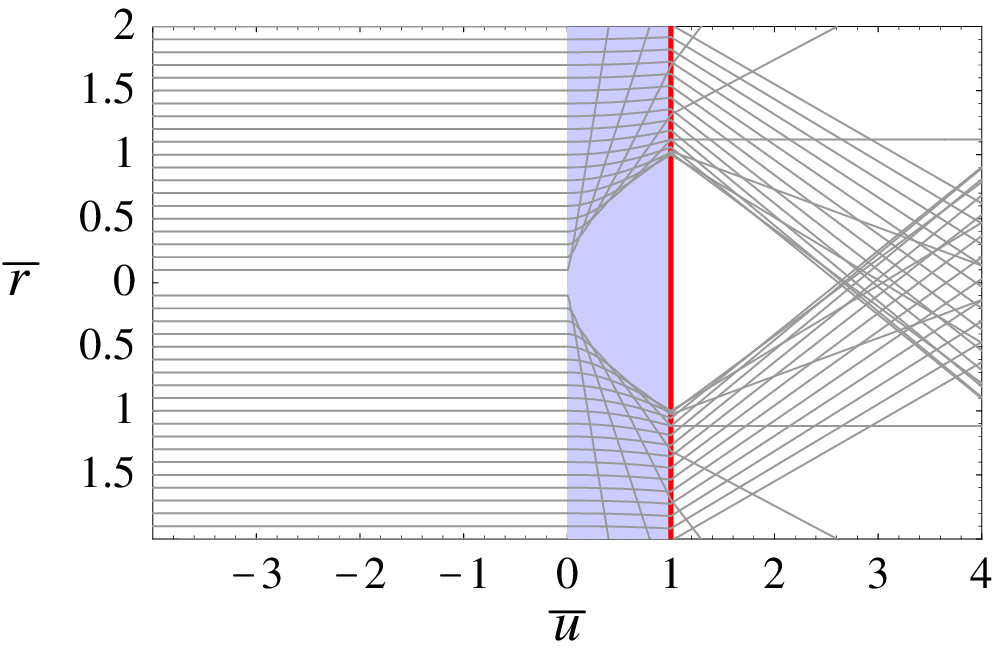}
}
\caption{The same as Fig.~\ref{nullray-advance} but for a {\it b}--gyraton.}
\label{nullray-follow}
\end{figure}
%

Finally we obtain
the formulas for $G$ and $H$ of a spinning {\it b}--gyraton.
They are found
by solving Eqs. \eqref{eq-Hu} and \eqref{equation-for-G} 
using Eq.~\eqref{chip-chij-typeb}. The result is
%
\begin{equation}
G(u,r)=
\begin{cases}
\tilde{G}(u,r),
&
(0\le u\le L),
\\
\displaystyle
\frac{(j^2-Lr^2)(u-L)}{Lr\sqrt{r^4+j^2}}+\tilde{G}(L,r),
&
(L\le u);
\end{cases}
\label{G-formula-typeb}
\end{equation}
%
%
\begin{equation}
H(u,r)=\begin{cases}
\tilde{H}(u,r),
&
(0\le u\le L),
\\
\tilde{H}(L,r),
&
(L\le u),
\end{cases}
\label{H-formula-typeb}
\end{equation}
%
where
%
\begin{equation}
\tilde{G}(u,r):=
r\sqrt{1+\frac{j^2/L^2}{(r^2/u)^2}};
\label{tildeG-formula-typeb}
\end{equation}
%
%
\begin{equation}
\tilde{H}(u,r):=
-\arctan \frac{j/L}{r^2/u}.
\label{tildeH-formula-typeb}
\end{equation}
%
Similarly to the {\it a}--gyraton, there are the focusing singularity at
%
\begin{equation}
u=\frac{r^2(r^2/L+1)}{r^2/L-j^2}, ~~(L\le u),
\end{equation}
%
and the crossing singularity at
%
\begin{equation}
u=\begin{cases}
(L/|j|) \ r^2,
&
(0\le u\le L),
\\
L-\tilde{G}_{,r}(L,r)/\tilde{G}_{,ur}(L,r),
&
(L\le u).
\end{cases}
\end{equation}
%
The shape of the two singularities in $(u,r)$ coordinates and
the propagation of light rays in $(\bar{u}, \bar{r})$ coordinates
are shown in Fig.~\ref{nullray-follow}.

%
%
\subsection{Gyraton collisions}

Consider two gyratons and assume that each of them
belongs to one of the four types described above.
We obtain several different configurations
for the collisions of these gyratons.
To illustrate main features of these collisions, 
in this subsection we set up five cases of the head-on collision 
of two gyratons which are of the most interest. 
The incoming gyratons are referred to as gyraton~1 and gyraton~2.
Let us divide a spacetime for the two-gyraton system into four regions:
%
\begin{equation}
\begin{array}{ll}
\textrm{Region I:} & (u\le 0, v\le 0),\\
\textrm{Region II:} & (u\ge 0, v\le 0),\\
\textrm{Region III:} & (u\le 0, v\ge 0),\\
\textrm{Region IV:} & (u\ge 0, v\ge 0).
\end{array}
\end{equation}
%
Because the gyratons propagate at the speed of light,
they do not interact with each other before the collision.
Thus we can use the metric of the gyraton~1 in the regions I and II 
and the metric of the gyraton~2 in the regions I and III
(by changing $u$ and $v$). 
Then,  the metric of the system is given as  
\begin{equation}
ds^2=\begin{cases}
-dudv+dr^2+r^2d\phi^2, 
& (\text{Region I}),
\\
-dudv+G^{(1)}_{,r}(u,r)^2dr^2+G^{(1)}(u,r)^2(d\phi+H^{(1)}_{,r}(u,r)dr)^2,
& (\text{Region II}),
\\
-dudv+G^{(2)}_{,r}(v,r)^2dr^2+G^{(2)}(v,r)^2(d\phi+H^{(2)}_{,r}(v,r)dr)^2, 
& (\text{Region III}).
\end{cases}
\end{equation}
%
The metric of the region IV is unknown, because the interaction
between the two gyratons determines its structure
through the Einstein equations.\footnote{
In Sec. IV, we will solve the Einstein equation in a part of the region 
IV in a specific case, when spins are small.}

%
\begin{figure}[tb]
\centering
{
\includegraphics[height=0.12\textheight]{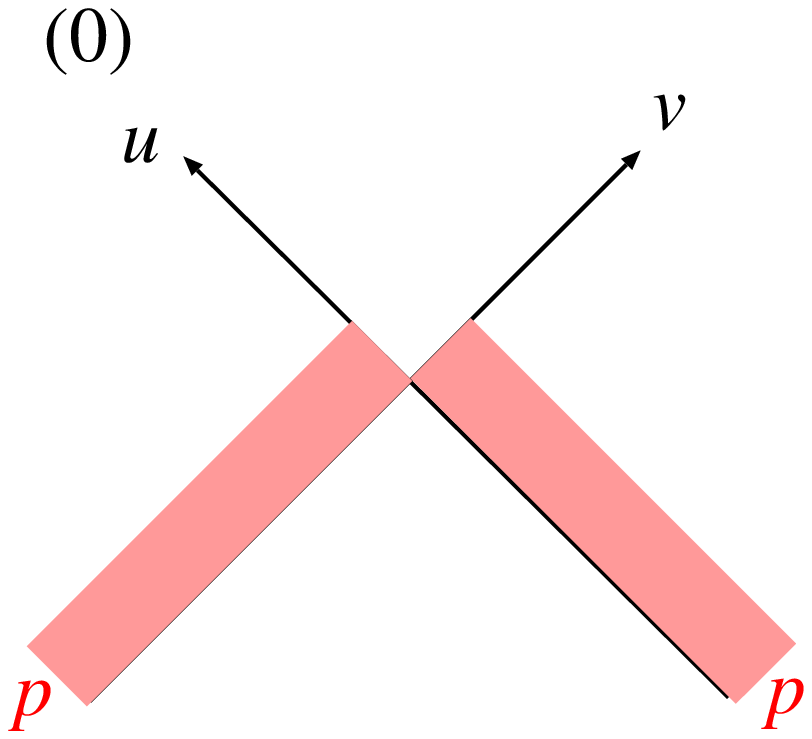}
\includegraphics[height=0.12\textheight]{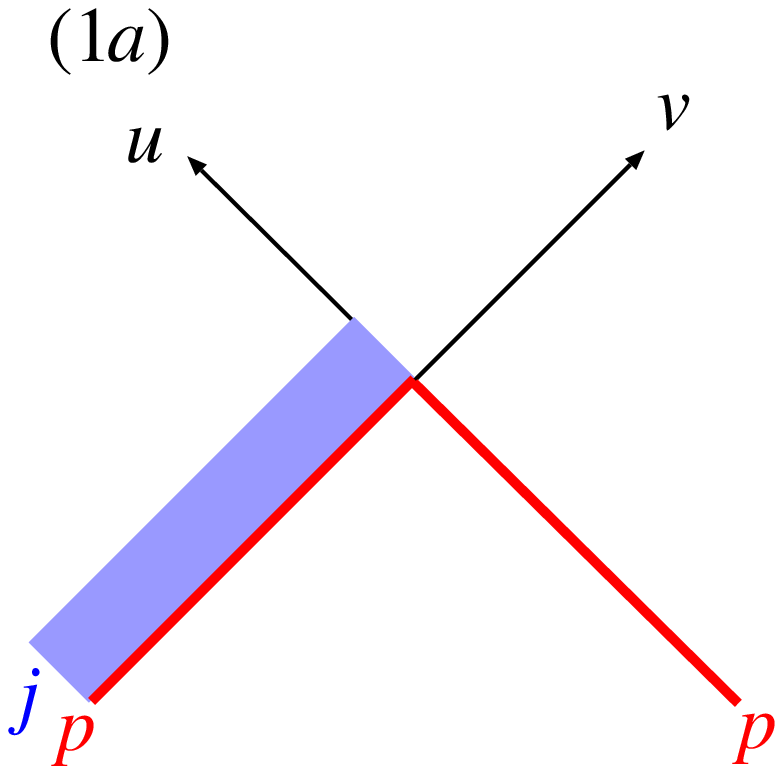}
\includegraphics[height=0.12\textheight]{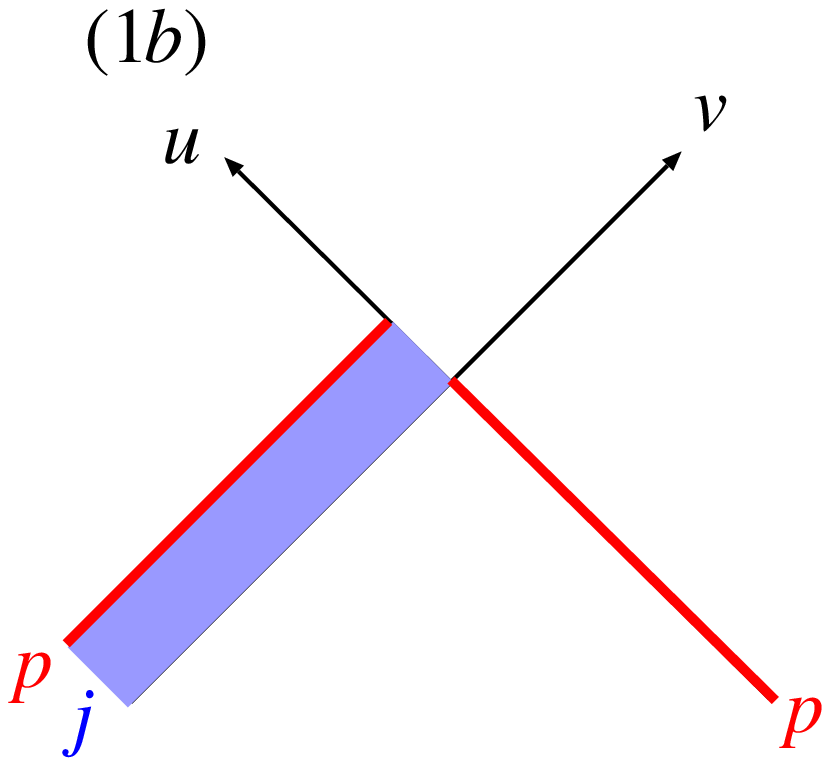}
\includegraphics[height=0.12\textheight]{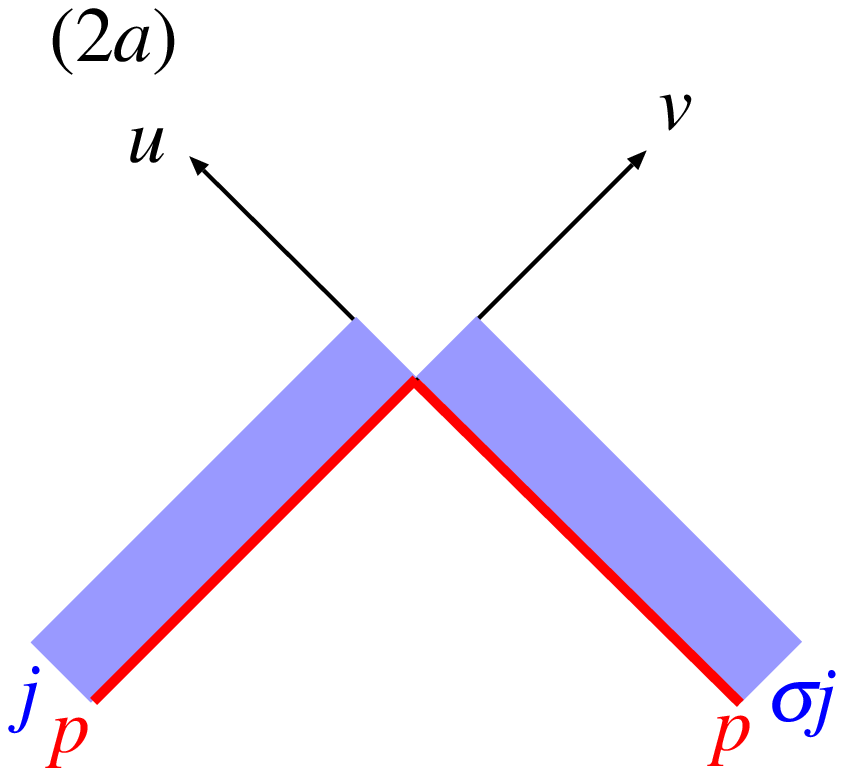}
\includegraphics[height=0.12\textheight]{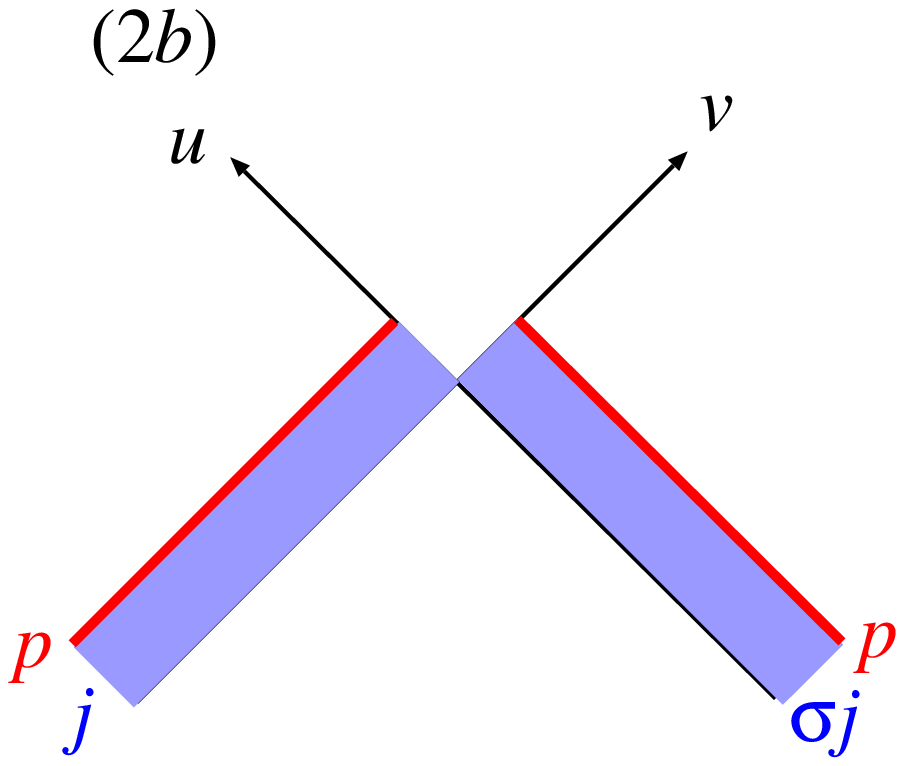}
}
\caption{The five cases of gyraton collision that we study
in this paper. In case (0), two identical $p$--gyratons collide.
In cases (1{\it a}) and (1{\it b}),  {\it a}-- and {\it b}--
gyratons with spin $j$ collide with an AS particle, respectively.
In cases (2{\it a}) and (2{\it b}), two {\it a}-- and {\it b}--
gyratons with spins $j$ and $\sigma j$ $(\sigma=\pm 1)$ collide, 
respectively.}
\label{collision-setup}
\end{figure}
%

In the previous subsection, we introduced four gyraton models, i.e.,
a $p$--gyraton, an AS--gyraton, an {\it a}--gyraton and 
a {\it b}--gyraton.
Using these models, we will consider five cases of collision.
The first one, which we call the case (0), is the collision of
two identical spinless $p$--gyratons. For both $G^{(1)}$ and $G^{(2)}$, 
we use the formula of $G$ for the $p$--gyraton 
\eqref{G-formula-withoutspin}--\eqref{def-y}.
The energy $p$ determins the scale, so that
the only essential parameter which specifies the system 
is the energy duration $L$.

In the next two cases (1{\it a}) and (1{\it b}), we consider collisions of 
a spinning gyraton ({\it a}-- and {\it b}-- gyraton, respectively) 
with an AS--gyraton. In both cases,
we assume that incoming gyratons have the same energy, 
and only a gyraton~1 has the spin $j$. 
These are interpreted as collisions of a spinning particle 
and a particle without spin.
For $G^{(1)}$ and $H^{(1)}$, 
we use the formulas 
\eqref{G-formula-typea}--\eqref{tildeH-formula-typea}
of $G$ and $H$ for {\it a}--gyraton in the (1{\it a}) case,  
and use the formulas
\eqref{G-formula-typeb}--\eqref{tildeH-formula-typeb}
of $G$ and $H$ for {\it b}--gyraton in the (1{\it b}) case.  
For $G^{(2)}$, the formula \eqref{G-formula-AS} of $G$ for AS--gyraton
is used in both cases.
The essential parameters which specify the system are the spin $j$ and
its duration $L$ for a gyraton~1.

In the remaining two cases (2{\it a}) and (2{\it b}), we study collisions
of two {\it a}--gyratons and two {\it b}-- gyratons, respectively.
These are interpreted as collisions of two spinning particles.
In these cases, the incoming gyratons are assumed 
to have the same energy $p$ and the same spin duration $L$.
As for the spin values, we assume that 
the gyraton~1 has the spin $j$ and 
the gyraton~2 has the spin $\sigma j$, where $\sigma=\pm 1$.
Therefore two spins have the same absolute value $|j|$
and have either the same sign or different signs.
For the choice $\sigma=+1$ two spins have the same direction (i.e.
helicities have opposite signs), and for the choice $\sigma=-1$
two spins have opposite directions (i.e. helicities have the same sign).
For $G^{(1)}$ and $H^{(1)}$,
we use the formulas
\eqref{G-formula-typea}--\eqref{tildeH-formula-typea}
of $G$ and $H$ for {\it a}--gyraton 
in the (2{\it a}) case, and
use the formulas
\eqref{G-formula-typeb}--\eqref{tildeH-formula-typeb}
of $G$ and $H$ for {\it b}--gyraton 
in the (2{\it b}) case.
For $G^{(2)}$ and $H^{(2)}$, 
we use the formulas of $G$ and $H$
for {\it a}-- and {\it b--} gyratons with $j$
replaced by $\sigma j$
in the (2{\it a}) and (2{\it b}) cases, respectively.
The essential parameters which specify the system 
are the spin $j$ of gyraton~1, the relative directions of two spins $\sigma$, 
and the spin duration $L$ of each incoming gyraton.
In the study of Secs. III--IV, the 
condition for AH formation in the slice $u=0\le v$ and $v=0\le u$
will turn out to be independent of $\sigma$, and hence
the essential parameters are reduced to $j$ and $L$. 
The sign of $\sigma$ will become important
in the study of the spin-spin interaction in Sec. V.

All the five cases are schematically illustrated in Fig.~\ref{collision-setup}.

%
%
\section{Finding the apparent horizon}

%
\begin{figure}[tb]
\centering
{
\includegraphics[width=0.5\textwidth]{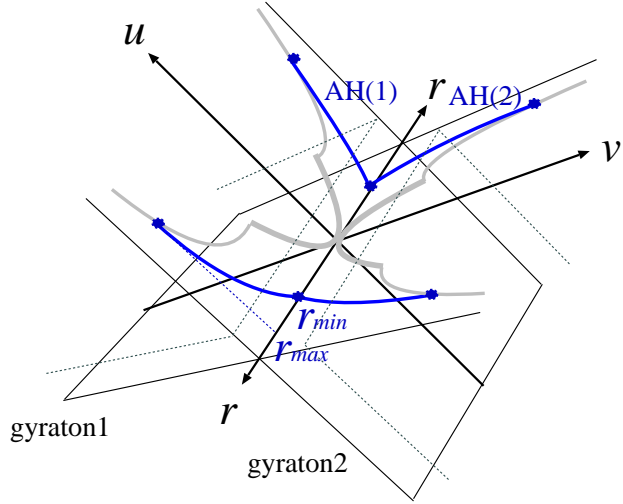}
}
\caption{The schematic picture of the AH in the slice $u\ge 0=v$ and 
$v\ge 0=u$.}
\label{AH-schematic-gyraton}
\end{figure}
%

The apparent horizon (AH) $\Sigma$ is a compact
two-dimensional spacelike surface for which the family of outgoing
null rays emitted orthogonally to $\Sigma$ has zero expansion.
We study the AH on the slice $u\ge 0=v$ and $v\ge 0=u$.
Figure \ref{AH-schematic-gyraton} shows a schematic picture of the AH
for colliding gyratons.
The AH consists of two parts:
%
\begin{equation}
\begin{cases}
u=h^{(1)}(r), & (u>0=v),\\
v=h^{(2)}(r), & (v>0=u).
\end{cases}
\label{ah-specify}
\end{equation}
%
These parts are connected at $u=v=0$. Each
surface has the other end at the focusing singularity.
Because the focusing singularity is just one point for the same $r$ value,
the surface given by Eq.~\eqref{ah-specify}
is a two-dimensional closed spacelike surface.

Because
the AH equations and the outer boundary conditions 
for $h^{(1)}$ and $h^{(2)}$ have the same form, we 
only consider $h(r):= h^{(1)}(r)$ and
denote $G=G^{(1)}$ and $H=H^{(1)}$. 
The metric in the neighborhood of $v=0<u$ is given by
%
\begin{equation}
ds^2=-dudv+G_{,r}^2dr^2+G^2(d\phi+H_{,r}dr)^2.
\end{equation}
%
Let us consider a point $(u,r,\phi)=(h(R),R,\varPhi)$. 
The local lightcone with the apex at this point is 
%
\begin{equation}
(u-h(R))v=G_{,r}^2(r-R)^2+G^2[(\phi-\varPhi)+H_{,r}(r-R)]^2.
\label{local-light-cone}
\end{equation}
%
We find the envelope of the local lightcones by 
taking the derivative of Eq.~\eqref{local-light-cone}
with respect to $R$ and $\varPhi$.
The tangent vector of the null geodesic congruence 
in the $(u,v,r,\phi)$ coordinate is
%
\begin{equation}
k^{\mu}=
\left(
\frac{h_{,r}^2}{2G_{,r}^2},
2,
\frac{h_{,r}}{G_{,r}^2},
-\frac{H_{,r}}{G_{,r}^2}h_{,r}
\right).
\end{equation}
%

Now we calculate the expansion.
The induced metric on $v=\text{const}$ surface
is given by
$d\gamma^2=G_{,r}^2dr^2+G^2(d\phi+H_{,r}dr)^2$,
and its determinant $\gamma$ is $\sqrt{\gamma}=G_{,r}G$.
Let us consider a 
rectangular coordinate domain with apices at
the four points $P_1$, $P_2$, $P_3$ and $P_4$:
%
\begin{equation}
(r,\phi)=
\left\{
\begin{array}{cc}
P_1: &(R_+, \varPhi_+),\\
P_2: &(R_+, \varPhi_-),\\
P_3: &(R_-, \varPhi_+),\\
P_4: &(R_-, \varPhi_-).
\end{array}
\right.
\end{equation}
%
Here $R_\pm=R \pm \Delta r/2$ and $\varPhi_\pm =\varPhi \pm \Delta\phi/2$.
The proper area $\Delta A(0)$ of this domain is
%
\begin{equation}
\Delta A(0)=\left.G_{,r}G\right|_{(h(R),R)}\Delta r \Delta\phi.
\end{equation}
%
The null geodesics passing through the apices are
%
\begin{equation}
(r,\phi)=
\left\{
\begin{array}{c}
P_1^\prime:~~
(R_+ + k^r(R_+)\lambda, 
\varPhi_+ + k^\phi(R_+)\lambda),
\\
P_2^\prime:~~
(R_+ + k^r(R_+)\lambda, 
\varPhi_- + k^\phi(R_+)\lambda),
\\
P_3^\prime:~~
(R_- + k^r(R_-)\lambda, 
\varPhi_+ + k^\phi(R_-)\lambda),
\\
P_4^\prime:~~
(R_- + k^r(R_-)\lambda, 
\varPhi_-+k^\phi(R_-)\lambda),
\end{array}
\right.
\end{equation} 
%
where $\lambda$ is an affine parameter.
In what follows, we keep 
terms up to first order in $\lambda$.
The coordinate shape of the domain
surrounded by the four apices after evolution 
is a parallelogram as indicated by the vectors
%
\begin{eqnarray}
\overrightarrow{P_2^\prime P_1^\prime}=
\overrightarrow{P_4^\prime P_3^\prime}&=&(0, \Delta\phi),\\
\overrightarrow{P_3^\prime P_1^\prime}=
\overrightarrow{P_4^\prime P_2^\prime}&=&
((1+\partial_rk^r\lambda)\Delta r, \partial_rk^\phi\lambda\Delta r).
\end{eqnarray}
%
The coordinate area of the domain is
$(1+\partial_rk^r\lambda)\Delta r\Delta\phi$
and the proper area of the domain is 
%
\begin{eqnarray}
\Delta A(\lambda)&=&
\left.\left(G_{,r}G\right)
\right|_{(h(R)+k^u(R)\lambda,R+k^r(R)\lambda)}
\left(1+\partial_rk^r\lambda\right) \Delta r\Delta\phi\nonumber\\
&=&\Delta A(0)
\left[
1+\left(
\frac{G_{,ru}k^u+G_{,rr}k^r}{G_{,r}}
+\frac{G_{,u}k^u+G_{,r}k^r}{G}
+\partial_rk^r
\right)\lambda
\right].
\end{eqnarray}
%
Hence, the condition $d\Delta A/d\lambda=0$ implies
%
\begin{equation}
h_{,rr}+h_{,r}
\left[
-\frac{(3/2)G_{,ru}h_{,r}+G_{,rr}}{G_{,r}}
+
\frac{(1/2)G_{,u}h_{,r}+G_{,r}}{G}
\right]=0.
\label{AH-equation}
\end{equation}
%
This equation determines the AH surface $\Sigma$.

Let us discuss now the boundary conditions at the outer boundary $r=r_{\rm max}$.
By the continuity of the surface, 
the AH should cross the coordinate singularity
at $r=r_{\rm max}$:
%
\begin{equation}
G(h(r_{\rm max}), r_{\rm max})=0.
\label{outer-bc-1}
\end{equation}
%
The continuity of $k^\mu$ also must be imposed,
because the surface has a delta-functional expansion 
if $k^{\mu}$ is discontinuous.
Going back to the $(\bar{u}, \bar{v}, \bar{r}, \bar{\phi})$
coordinate, the continuity can be imposed as
$k^{\bar{r}}=0$
by the axisymmetry. This is equivalent to
%
\begin{equation}
h_{,r}(r_{\rm max})=-2{G_{,r}}/{G_{,u}}.
\label{outer-bc-2}
\end{equation}
%
The other condition for the continuity of $k^\mu$
is that $k^{\bar{\phi}}$ should take a finite value. 
But this is automatically satisfied 
since the condition $H_{,u}=0$ at the focusing singularity
implies that $k^{\bar{\phi}}=0$.

Now we turn to the boundary conditions
at the inner boundary $r=r_{\mathrm{min}}$. 
The inner boundary conditions depend on both $h^{(1)}$ and $h^{(2)}$.
By continuity of the surface,
both sides of the AH should cross $u=v=0$ at $r=r_{\rm min}$, and thus
%
\begin{equation}
h^{(1)}(r_{\rm min})=h^{(2)}(r_{\rm min})=0.
\label{inner-bc-1}
\end{equation}
%
Also the null tangent vectors ${k^{(1)}}^\mu$
and ${k^{(2)}}^\mu$ of two surfaces should be parallel 
at $r=r_{\mathrm{min}}$ so that
there is no delta-functional expansion.
${k^{(1)}}^\mu$ and ${k^{(2)}}^\mu$ are given by
%
\begin{equation}
{k^{(1)}}^\mu
= \left( 2, \frac{{h^{(1)}_{,r}}^2}{2}, h^{(1)}_{,r}, 0 \right),
\
{k^{(2)}}^\mu
= \left( \frac{{h^{(2)}_{,r}}^2}{2}, 2, h^{(2)}_{,r}, 0 \right),
\end{equation}
%
and $k^{(1)}\parallel k^{(2)}$ is equivalent to
%
\begin{equation}
h^{(1)}_{.r}(r_{\rm min})h^{(2)}_{,r}(r_{\rm min})=4.
\label{inner-bc-2}
\end{equation}
%

The numerical procedure
for defining the AH is straightforward. First we choose some value
of $r_{\rm max}$ and solve $h(r)$ with the outer boundary conditions
\eqref{outer-bc-1} and \eqref{outer-bc-2}.
When $h(r)$ becomes zero at $r=r_{\rm min}$, we check whether the
inner boundary condition \eqref{inner-bc-2} is satisfied.
Iterating these steps for various values of $r_{\rm max}$, we 
determine whether the AH exists and find its location.

Note that $H$ does not appear in the AH equation and the boundary 
conditions. This
means that the dragging-into-rotation effect 
causes a change in the shear but not in the expansion.
Thus on the slice we have adopted, the condition for the AH formation
does not depend on the sign of $j$
in the cases (1{\it a}) and (1{\it b}). 
In cases (2{\it a}) and (2{\it b}),
it does not depend also on the relative directions of two spins $\sigma$.
Therefore in the next section $j$ is 
assumed to be positive without loss of generality
and we do not specify the value of $\sigma$ 
when the numerical results are shown.

%
%
\section{Numerical Results}

In this section, we present the numerical results of the AH studies.
The results for 
case (0), 
cases (1{\it a}) and (1{\it b}), 
and cases (2{\it a}) and (2{\it b}) 
are provided in Secs. IV A, IV B, and IV C, respectively.

%
%
\subsection{Collision of gyratons without spin}

%
\begin{figure}[tb]
 \centering 
 \includegraphics[width=0.2\textwidth]{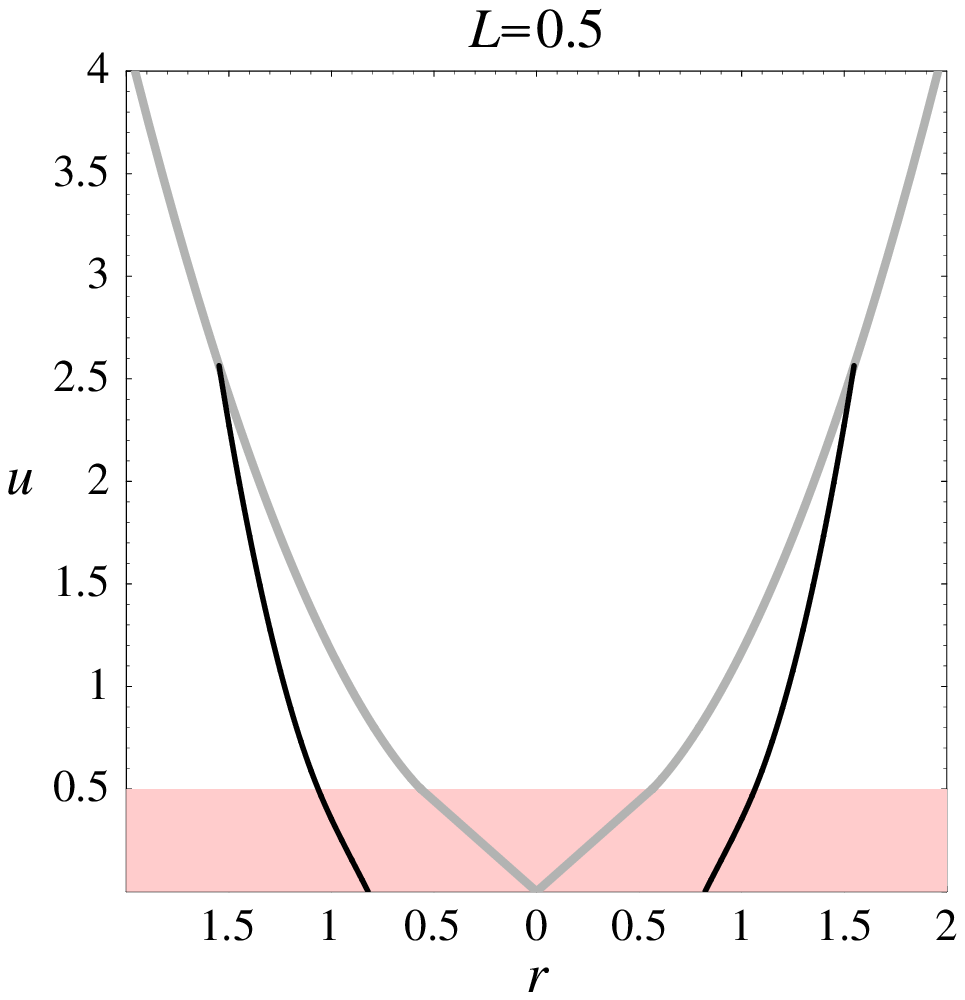}
 \includegraphics[width=0.2\textwidth]{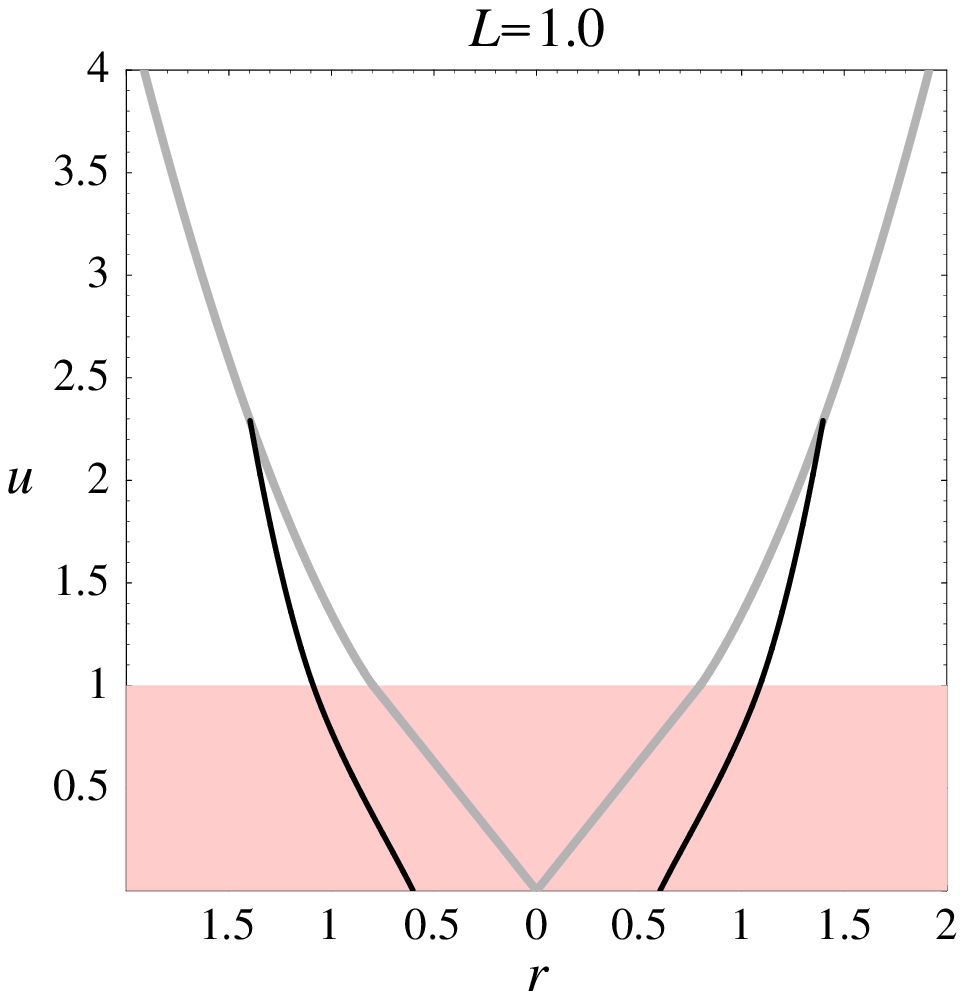}
 \includegraphics[width=0.2\textwidth]{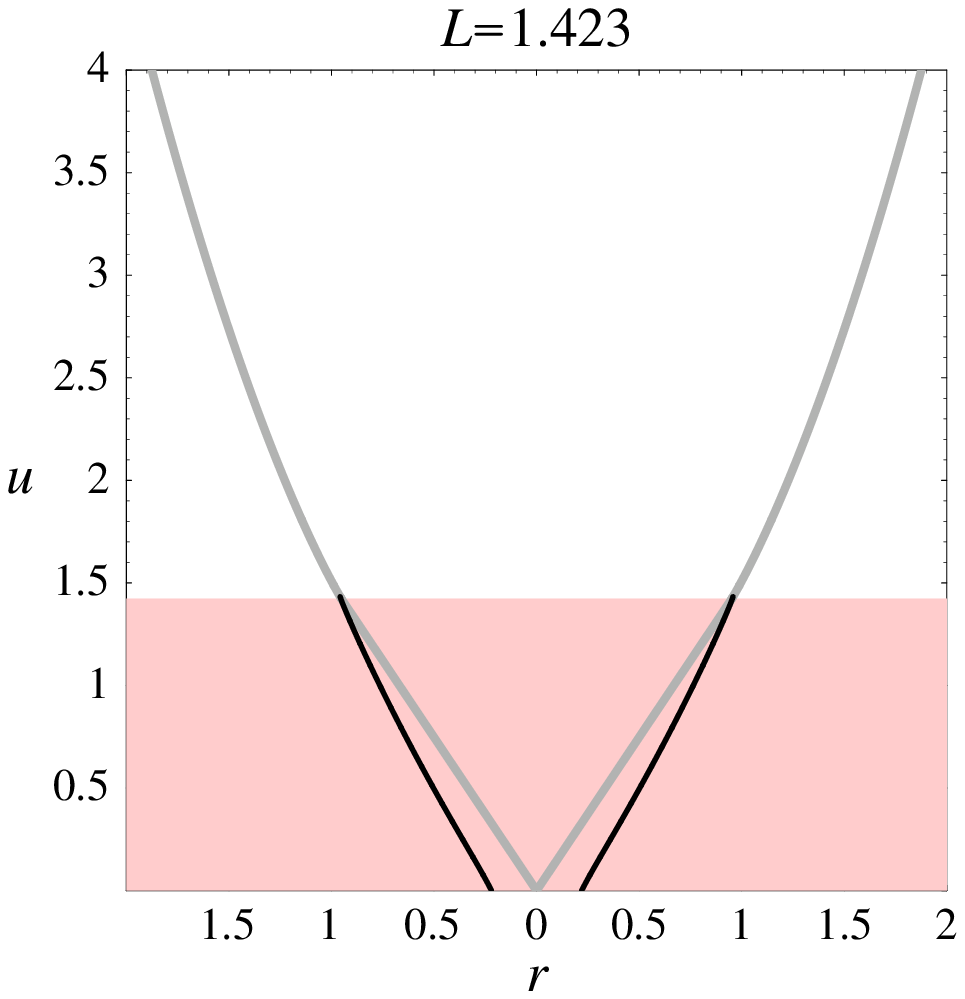}
  \caption{The top views of the AH in case
  (0) for $L=0.5, 1.0$, and $1.423$. For all $0\le L\le 1.423$, we find only one solution.
   At $L=1.423$, the AH almost touches
  the energy source at the symmetry axis. No AH exists for $L\ge 1.424$.
  }
 \label{AH-case0}
\end{figure}
%

We begin with case (0), the collision of two identical spinless $p$--gyratons
with energy duration $L$. 
Figure~\ref{AH-case0} shows top views of the AH 
for $L=0.5, 1.0$ and $1.423$. For $0\le L\le 1.423$, we found
only one solution, and therefore there is no inner boundary
of the trapped region. For $L=1.423$, the AH intersects
the focusing singularity at $u\simeq L$ and almost touches
the source of the energy which distributes at $0\le \bar{u}\le L$ on the symmetry axis.
For $L\ge 1.424$, we found no solution.
Thus, on the slice we studied, the condition of AH formation 
is given by $L\le 1.423$ in the length unit $r_h(2p)=1$.

%
%
\subsection{Collision of a spinning gyraton with an AS particle}

Next we show the results of cases (1{\it a}) and (1{\it b}), 
i.e., the collisions of spinning {\it a}-- and {\it b}-- gyratons 
with an AS particle.

%
\begin{figure}[tb]
 \centering 
 \includegraphics[width=0.2\textwidth]{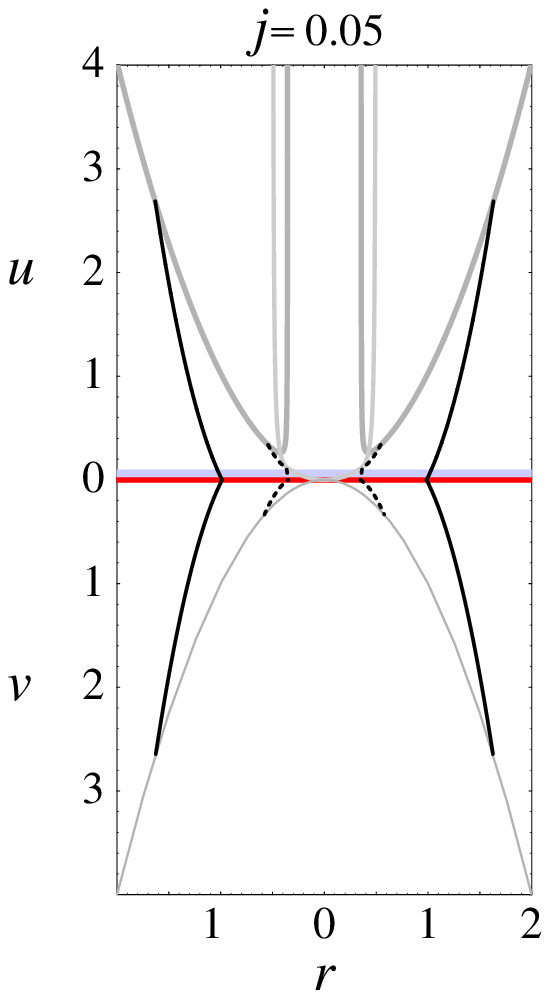}
 \includegraphics[width=0.2\textwidth]{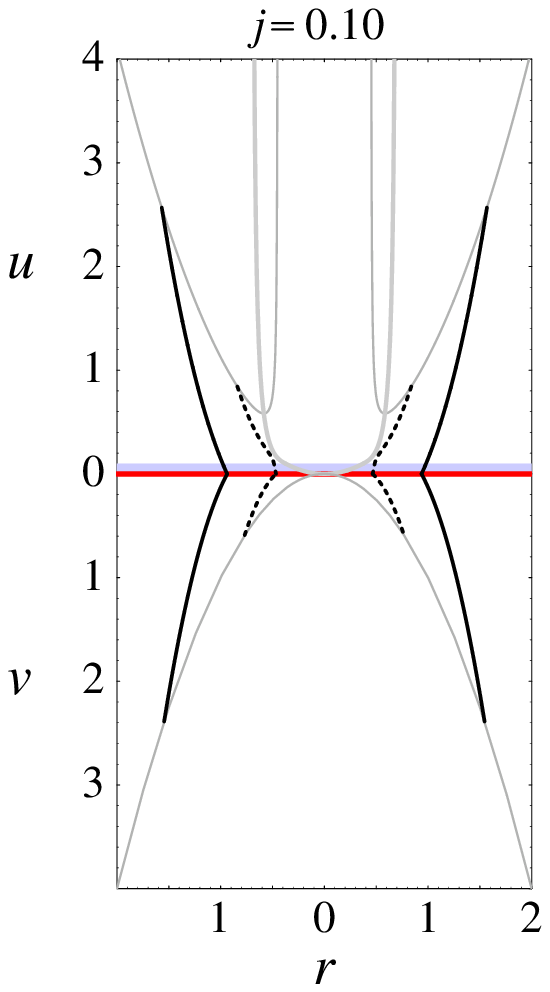}
 \includegraphics[width=0.2\textwidth]{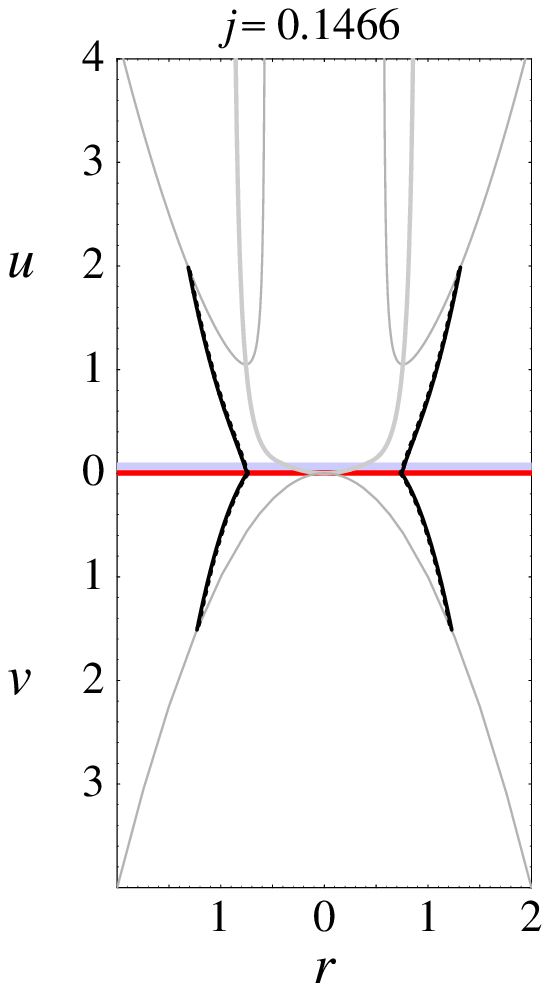}
   \caption{The top views of the AH in case
  (1{\it a}) for $L=0.1$ and $j=0.05,0.10$, and $j_{\rm crit}=0.1466$. 
  For all $0< j\le 1.466$, there are
  the AH (solid lines) and the inner boundary of the trapped region (dashed lines).
  The trapped region shrinks as $j$ is increased 
  and no AH exists for $j\ge 0.1467$.
  }
 \label{AH-case1a}
\end{figure}
%

%
\begin{figure}[tb]
 \centering 
 \includegraphics[width=0.2\textwidth]{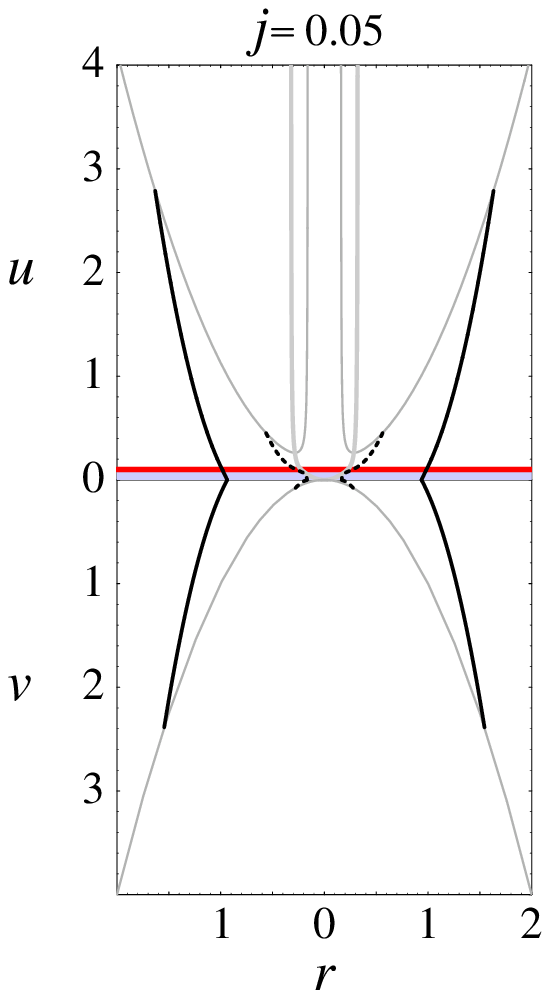}
 \includegraphics[width=0.2\textwidth]{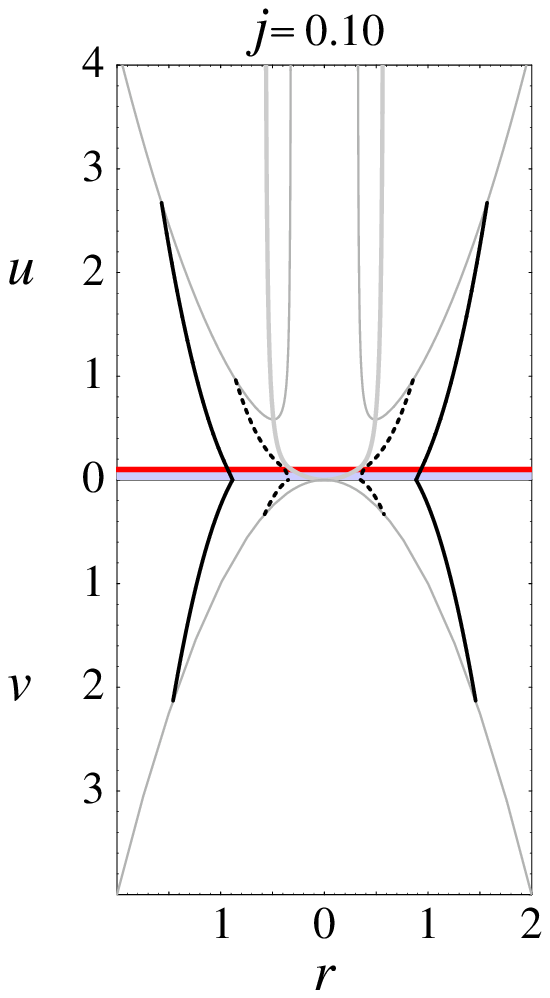}
 \includegraphics[width=0.2\textwidth]{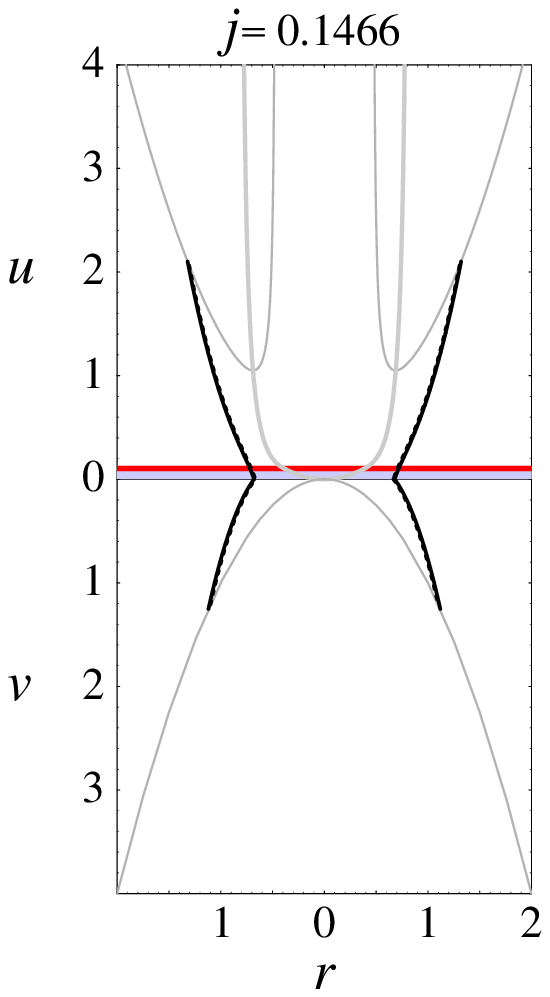}
   \caption{The same as Fig.~\ref{AH-case1a} but for case (1{\it b}).
   For $L=0.1$, the critical value of AH formation $j_{\mathrm{crit}}$ 
   is almost the same as that in case (1{\it a}).
  }
 \label{AH-case1b}
\end{figure}
%

Figure~\ref{AH-case1a} shows the AHs in case (1{\it a}) for
parameters $L=0.1$ and $j=0.05,0.10$, and $0.1466$. 
We found two solutions to the AH equation, which correspond to
the AH and the inner boundary of the trapped region.
As the value of $j$ is increased for a fixed value of $L$,
the trapped region grows smaller and the two solutions coincide
at some value of $j=j_{\rm crit}$. 
The trapped region vanishes for $j\ge j_{\rm crit}$.
The similar phenomena was observed also in case (1{\it b}).
In Fig.~\ref{AH-case1b}, the AH shape in case (1{\it b})
is shown for the same parameters as those in Fig.~\ref{AH-case1a}.
Again, there are two solutions and they degenerate at some critical value
$j=j_{\rm crit}$. 

As it has been found above, 
the spin makes the formation of the AH more ``difficult''.
This is because the gravitational field generated
by the spin source is repulsive
as we pointed out in Sec. II. 
As the value of $j$ is increased,
the repulsive force surpasses the attractive
force generated by the energy source and causes the extinction of the AH.

%
\begin{figure}[tb]
 \centering 
 \includegraphics[width=0.4\textwidth]{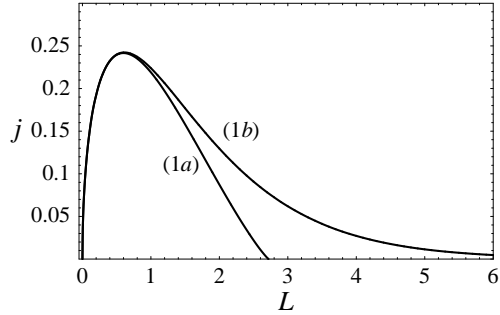}
 \caption{The critical line on $(L,j)$-plane for AH formation in the cases (1{\it a}) and (1{\it b}). 
 The AH formation is allowed under the critical line. 
 The two critical lines almost coincide for $L\le 1$ and go to zero in the limit $L\to 0$.
 The critical line of the (1{\it a}) case intersects the $L$-axis at $L=e$, while
 that of the (1{\it b}) case becomes exponentially close to the $L$-axis 
 as $L$ is increased.
 }
 \label{region-case1}
\end{figure}
%

We studied the value of $j_{\rm crit}$ as a function of $L$, 
i.e., $j_{\rm crit}(L)$.
In Fig.~\ref{region-case1},
the critical lines for the AH formation in the $(L,j)$-plane 
are shown for both cases. The AH formation is allowed
under the critical line. For $L\lesssim 1$, the two critical lines agree well
and go to zero in the limit $L\to 0$. Both critical lines have the peak
$j\simeq 0.24$ at $L\simeq 0.6$.
For $L\gtrsim 1$, the two critical lines show different behaviors.
In case (1{\it a}), $j_{\rm crit}(L)$ decreases and becomes zero at $L=e:=\exp(1)$.
On the other hand, it decays (almost) exponentially
but never becomes zero in case (1{\it b}).

Let us discuss the reasoning for these results.
The reason why $j_{\rm crit}$ goes to zero 
in the limit $L \to 0$ is as follows.
As stated above, the extinction of the AH is caused
by the repulsive force generated by the spin source.
Thus, it is useful to introduce a radius $r_{\rm eq}$ 
where the attractive force due to the energy and the repulsive
force due to the spin balance one another.
For this purpose, let us recall Figs.~\ref{nullray-advance} and \ref{nullray-follow}
that show the propagation of light rays through the gravitational field
of the gyratons. The figures indicate the existence of $r_{\rm eq}$
such that the rays with $r>{r_{\rm eq}}$ shrink and those with $r<r_{\rm eq}$
expand in the region $\bar{u}>L$. Such $r_{\rm eq}$ is found by the equation
$G_{,u}(L,r_{\rm eq})=0$ and solved as
%
\begin{equation}
r_{\mathrm{eq}}=
\begin{cases}
\sqrt{L+{j^2}/{L}}, & \textrm{{\it a}--gyraton}, \\
\sqrt{{j^2}/{L}}, & \textrm{{\it b}--gyraton}. \\
\end{cases}
\end{equation}
%
As $L$ is decreased, $r_{\rm eq}$ becomes
larger, which indicates that the repusive force becomes stronger.
It is natural that $r_{\rm eq}\lesssim 1$ represents the 
condition for AH formation, and it is reduced to $j^2\lesssim L$
in the limit $L\to 0$. This explains the behavior of the critical lines
at $L\ll 1$.

At $L\gtrsim 1$, the condition $r_{\rm eq}\lesssim 1$ does not 
explain the numerical results well. This is because the above
discussion takes account only of the gravitational structure in the
transverse direction of motion, which
would be sufficient in the case $L\ll 1$, while
the spin duration $L$
plays an important role for the AH existence
in the case $L\gtrsim 1$.  
Let us first consider case (1{\it a}).
Taking a limit $j\to 0$ for $L\gtrsim 1$,
the AH solution is expected to reduce to
that for the collision of two AS particles:
%
\begin{equation}
h^{(1)}(r)=h^{(2)}(r)=2r^2\log r,
\label{AH-solution-2AS}
\end{equation}
%
with $r_{\rm min}=1$ and $r_{\rm max}=\sqrt{e}$.
This statement holds only for $L<e$.
In the case $L>e$, the ``solution'' \eqref{AH-solution-2AS}
plunges into the crossing singularity. In other words, it
crosses the spin source distributed on the symmetry axis
for $0\le u\le L$, on which the outer boundary condition 
cannot be imposed. 
Thus in case (1{\it a}), the situations $j=0$ and $j=0_+$
are different. This is the reason why the critical line intersects the
$L$-axis at $L=e$.

Next we discuss case (1{\it b}). In the limit $j\to 0$
for $L\gtrsim 1$, the AH solution reduces to
%
\begin{equation}
h^{(1)}(r)=\begin{cases}
2r^2\log r+L, & (1\le r\le \sqrt{e}),\\
2\log(r/r_{\rm min}), & (r_{\rm min}\le r\le 1),
\end{cases}
\end{equation}
%
%
\begin{equation}
h^{(2)}(r)=2r^2\log(r/r_{\rm min}),~~(r_{\rm min}\le r\le \sqrt{e}r_{\rm min}),
\end{equation}
%
where $r_{\rm min}=e^{-L/2}$.
In contrast to the (1{\it a}) case, this statement is valid for arbitrary $L$, because
the AH never touches the spin source.
Then, the condition for AH formation in the case $j>0$ is expected to be 
$r_{\rm min}\gtrsim r_{\rm eq}$, which is equivalent to
$j^2\lesssim Le^{-L}$. This explains the exponential decay of the
critical line in the (1{\it b}) case.

Note that the above interpretations, especially the ones for $L\gtrsim 1$,
strongly depend on the slice we have adopted. Thus there is no reason
why the above discussion holds for another slice that is future to 
our slice. Hence, we should keep in mind the possibility that
the critical line does not touch the $L$-axis in another slice
also in the case (1{\it a}).

To summarize, 
for the collision of a spinning gyraton with the AS-gyraton
and for the slice we have adopted,
the condition for the AH formation is roughly expressed
as $L\sim 1$ and $j\lesssim 0.25$ in both cases.

%
%
\subsection{Collision of two spinning gyratons}

Finally we show the results of cases (2{\it a}) and (2{\it b}), 
i.e., the collisions of two spinning $a$-- and $b$-- gyratons
(identical up to helicities).

%
\begin{figure}[tb]
 \centering 
 \includegraphics[width=0.2\textwidth]{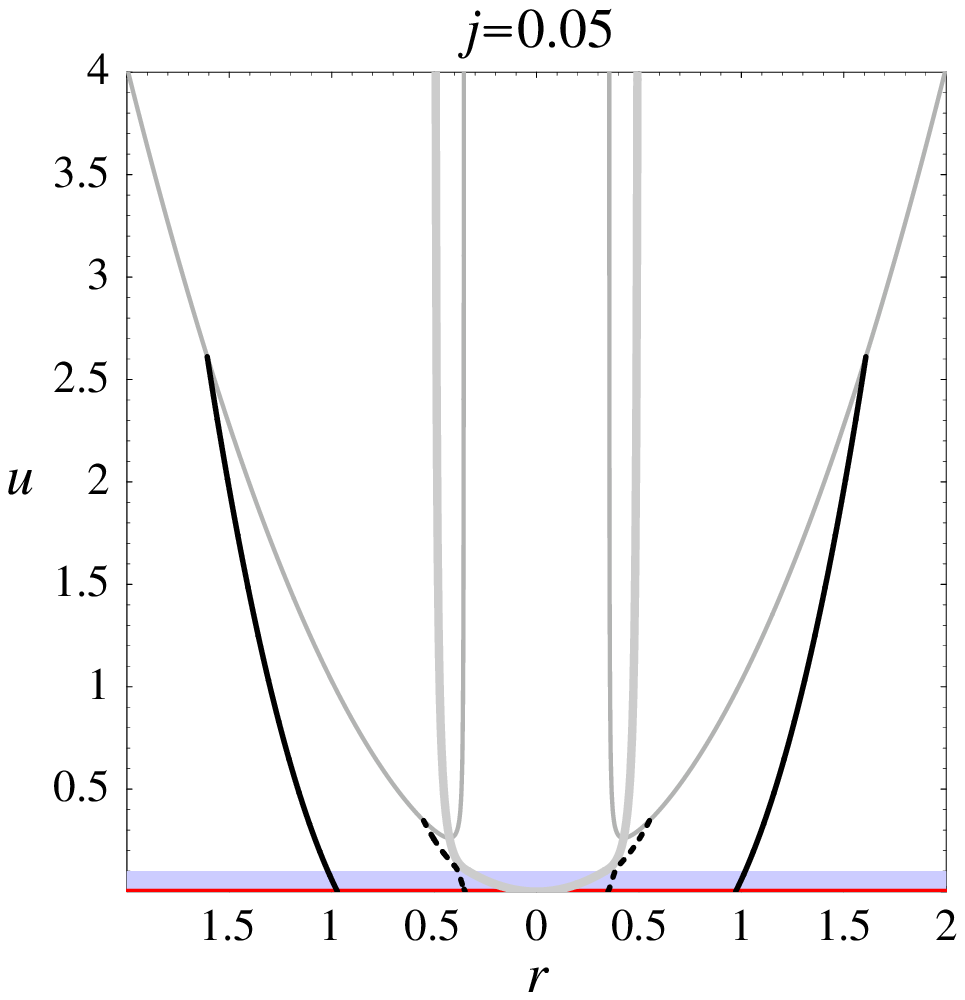}
 \includegraphics[width=0.2\textwidth]{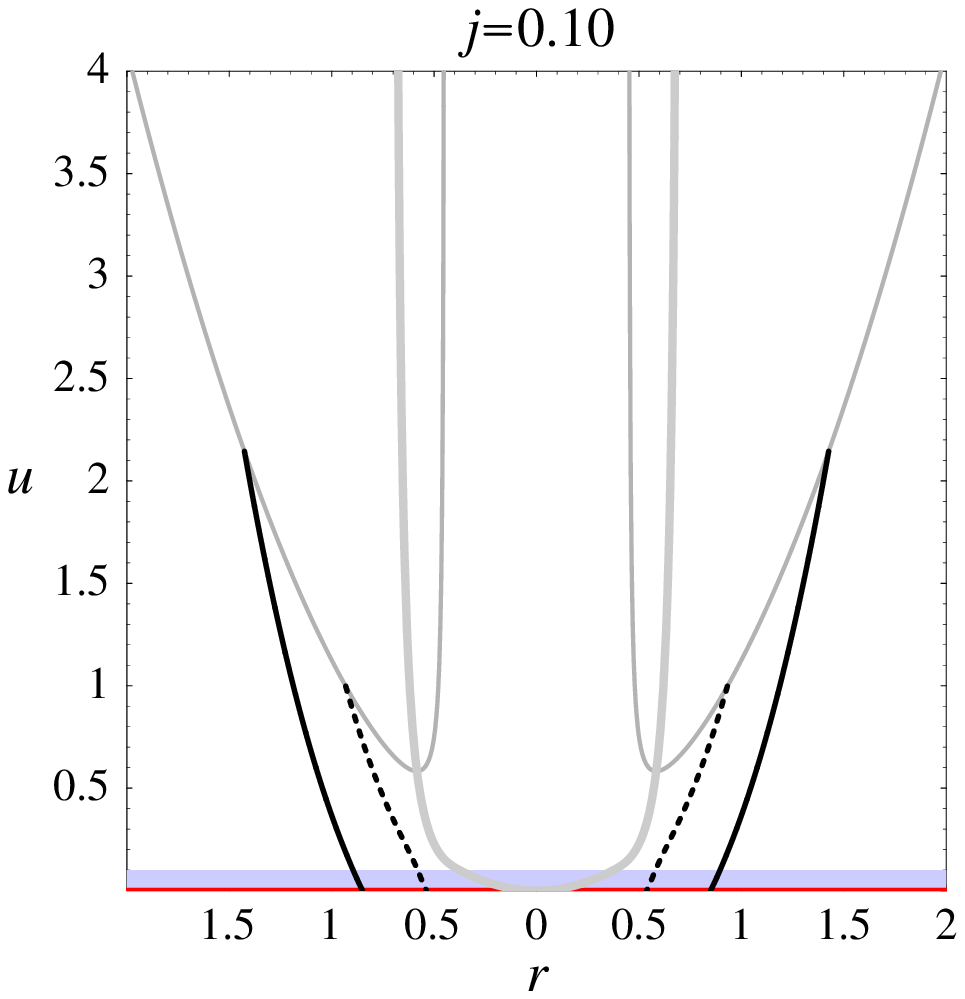}
 \includegraphics[width=0.2\textwidth]{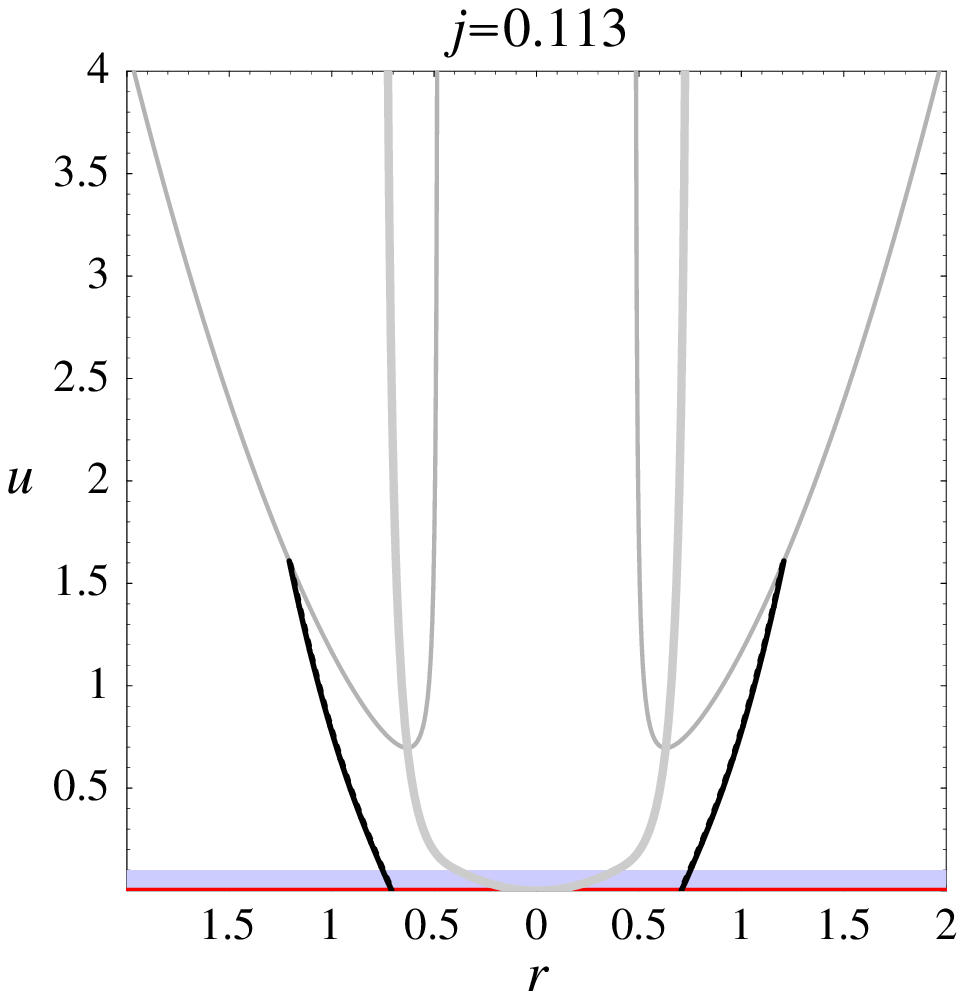}
   \caption{The top views of the AH (solid lines) and the inner boundary
   of the trapped region (dashed lines) in case
  (2{\it a}) for $L=0.1$ and $j=0.05,0.10$, and $0.113$. 
  No AH exists for $j\ge 0.113$.
  }
 \label{AH-case2a}
\end{figure}
%

%
\begin{figure}[tb]
 \centering 
 \includegraphics[width=0.2\textwidth]{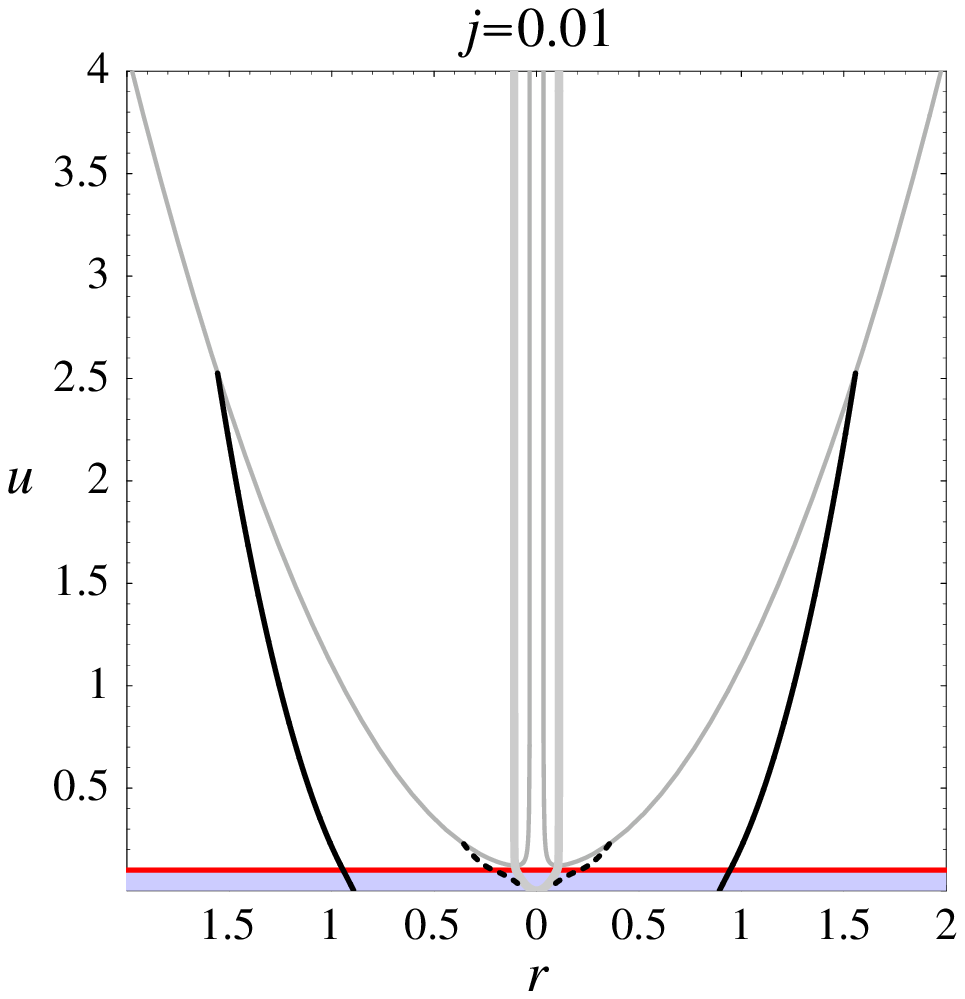}
 \includegraphics[width=0.2\textwidth]{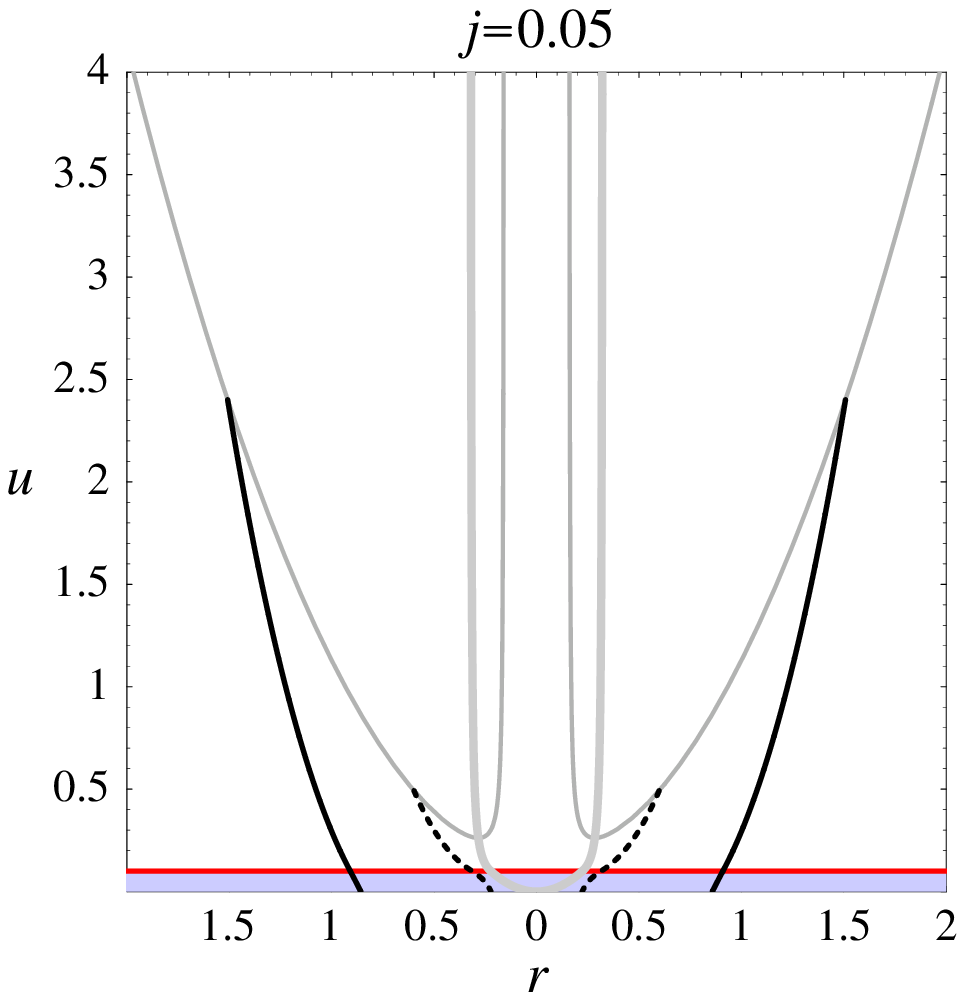}
 \includegraphics[width=0.2\textwidth]{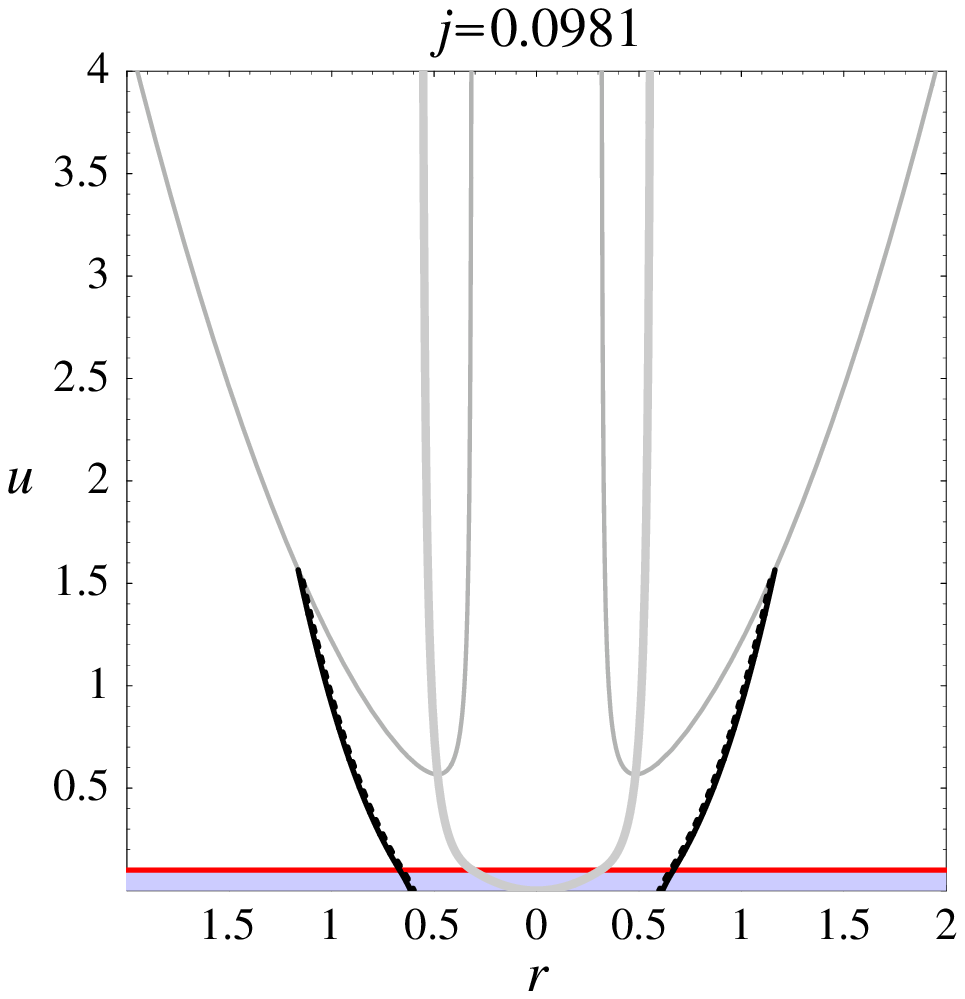}
   \caption{The top views of the AH (solid lines) and the inner boundary
   of the trapped region (dashed lines) in case
  (2{\it b}) for $L=0.1$ and $j=0.01,0.05$, and $0.0981$. 
  No AH exists for $j\ge 0.0982$.}
 \label{AH-case2b}
\end{figure}
%

Figure~\ref{AH-case2a} shows the AHs in case (2{\it a}) for
parameters $L=0.1$ and $j=0.05,0.10$, and $0.113$. 
Similarly to (1{\it a}) case,
there are two solutions to the AH equation, which 
surround the trapped region, and
they coincide at some value of $j=j_{\rm crit}$
as the value of $j$ is increased for a fixed value of $L$.                  
The similar phenomena was observed also in case (2{\it b}).
In Fig.~\ref{AH-case2b}, the AH shape in case (2{\it b})
is shown for $L=0.1$ and $j=0.01,0.05$, and $0.0981$.
Again, we found two solutions and their disappearance at some critical value
$j=j_{\rm crit}$. 
Similarly to cases
(1{\it a}) and (1{\it b}), the spin has the effect to 
make the AH formation more difficult.

%
\begin{figure}[tb]
 \centering 
 \includegraphics[width=0.4\textwidth]{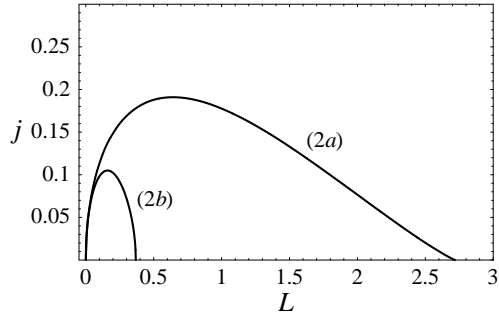}
 \caption{The critical lines on $(L,j)$-plane for AH formation in the cases (2{\it a}) and (2{\it b}). 
 The two lines go to zero in the limit $L\to 0$.
 The critical lines of (2{\it a}) and (2{\it b}) cases intersect the $L$-axis at $L=e$
 and $1/e$, respectively. }
 \label{region-case2}
\end{figure}
%

We studied the value of $j_{\rm crit}$ as a function of $L$.
In Fig.~\ref{region-case2},
the critical line for cases (2{\it a}) and (2{\it b}) in the $(L,j)$-plane 
is shown. The AH formation is allowed
under each critical line. 

Let us first discuss the critical line of the (2{\it a}) case.
It goes to zero in the limit $L\to 0$
and intersects the $L$-axis at $L=e$.
It has a peak
$j\simeq 0.19$ at $L\simeq 0.6$,  
and this peak value 
is somewhat smaller than the peak value $0.24$ of the (1{\it a}) case.
Hence the critical line of the (2{\it a}) case has the same features
as that of the (1{\it a}) case except that the peak value is smaller.
For the behaviors at $L\to 0$ and $L\to e$, 
the same reasoning to the results of the (1{\it a}) case holds.
Compared to the (1{\it a}) case, the AH formation 
is expected to become more difficult,
since both gyratons have the repulsive
forces around their centers in the (2{\it a}) case while only one gyraton
has the repulsive force in the (1{\it a}) case. This leads to the
smaller peak value of $j_{\rm crit}(L)$ in the (2{\it a}) case.

Now we discuss the critical line of the (2{\it b}) case. 
It goes to zero in the limit $L\to 0$
with the same reason to the (1{\it b}) case.
It has a peak
$j\simeq 0.105$ at $L\simeq 0.16$, and
intersects the $L$-axis at $L=1/e$. 
The allowed region of the (2{\it b}) case
is much smaller than that of the (2{\it a}) case.
The condition of the AH formation
strongly depends on the relative locations of the energy and spin profiles.
The reason can be understood as follows. In the limit $j\to 0$,
the AH becomes
%
\begin{equation}
h(r)=
\begin{cases}
2r^2\log\left(r/\sqrt{r_{\rm min}}\right)+L,
&
(\sqrt{r_{\rm min}}\le r\le \sqrt{er_{\rm min}}),
\\
2r_{\rm min}\log(r/r_{\rm min}),
&
(r_{\rm min}\le r\le \sqrt{r_{\rm min}}),
\end{cases}
\end{equation}
%
where $r_{\rm min}$ is given by the equation
%
\begin{equation}
L=-r_{\rm min}\log r_{\rm min}.
\end{equation}
%
This equation has two solutions for $0\le L<1/e$, one degenerate
solution for $L=1/e$, and no solution for $L>1/e$.
Thus the AH formation in the $j\to 0$ limit is allowed only
for $0\le L\le 1/e$. This is the reason why 
the allowed region is restricted to $0\le L\le 1/e$
and is much smaller than that of the case (2{\it a}).
However, we should keep in mind that this discussion 
is specific to the slice we have adopted. 
In the case $j=0$, the AH formation is allowed on a slice
appropriately taken at the future to our slice.
Hence, the allowed region in the (2{\it b}) case
is so small because of
the artificial effect of the slice choice.
In the next section, we demonstrate that this expectation
is true by solving a part of the spacetime after the collision
using the method of perturbation.

To briefly summarize, 
for the collision of two spinning gyratons,
the condition of the AH formation
on the slice we have adopted is roughly
expressed as $L\sim 1$ and $j\lesssim 0.2$
in the (2{\it a}) case, while $L\sim 0.15$ and $j\lesssim 0.1$ in the (2{\it b}) case.

%
%
\section{Second-order evolution}

In a general case, finding the spacetime structure
after the collision of two gyratons requires numerical
simulations. However, in the (2{\it b}) case, we
can go a little bit further using the method of perturbation
assuming that the spins of incoming gyratons are small.

%
\begin{figure}[tb]
 \centering 
  \includegraphics[width=0.4\textwidth]{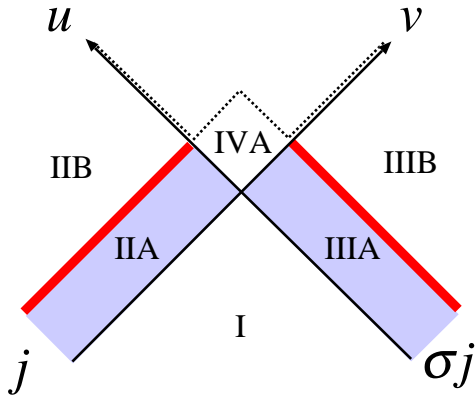}
 \caption{Schematic spacetime structure
 of the gyraton collision in the (2{\it b}) case. Region IVA can
 be solved using a perturbative method, assuming the spins are small. 
 Then we study the AH on the new slice shown by a dotted line, 
 which is the future edge of the solved region.}
 \label{region-IVA}
\end{figure}
%

Figure~\ref{region-IVA} shows the schematic spacetime structure
in the (2{\it b}) case where the gyraton 1 with the spin $j$ and
the gyraton 2 with the spin $\sigma j$ collide.
(Here $\sigma=\pm 1$, but we note that 
the solution to the Einstein equations
found in the Sec. IV A 
can be applied to an arbitrary value of $\sigma$.)
For $\sigma>0$ the two spins have the same direction
(i.e., the helicities of incoming gyratons have opposite signs),
while for $\sigma<0$ the two spins have opposite directions
(i.e., the helicities of incoming gyratons have the same sign).
The exact metrics in the regions I, IIA, IIB, IIIA and IIIB are known.
We focus our attention on finding the metric in the region IVA 
($0\le u\le L$ and $0\le v\le L$), where
the gravitational spin-spin interaction begins.
If the value of $j$ is small, we can expand the metric
in terms of $j$. 
The background spacetime is the Minkowski spacetime,
because the regions I, IIA, and IIIA
are flat for $j=0$.  
The first-order perturbation is easily solved. 
Because the metric in the regions IIA and IIIA
is
%
\begin{align}
ds^2=\begin{cases}
-dudv+dr^2+r^2d\phi^2+2\epsilon (u/r)drd\phi,&\textrm{region IIA},\\
-dudv+dr^2+r^2d\phi^2+2\epsilon (\sigma v/r)drd\phi,&\textrm{region IIIA},
\end{cases}
\end{align} 
%
where 
%
\begin{equation}
\epsilon=2j/L,
\end{equation}
%
the metric in the region IVA is found to be
%
\begin{equation}
ds^2=-dudv+dr^2+r^2d\phi^2+2\epsilon\frac{u+\sigma v}{r}drd\phi,
\label{first-order-solution}
\end{equation}
%
using the linearity of the first-order perturbation.
Strictly speaking, 
we have to specify the properties of matter interaction
between the sources of two incoming gyratons 
in order to determine the metric of the whole region IVA.
The domain where the matter interaction is important
is within the lightcone of the source collision, i.e. $uv>r^2$. 
The spacetime structure of the other domain $uv<r^2$ in the region IVA
is not affected by the matter interaction 
and therefore the metric~\eqref{first-order-solution}
can be applied for this domain. In the following, we restrict our attention
to the domain $uv<r^2$ and do not consider the effect
of matter interaction.

In order to study the  condition of AH formation,
the above first-order solution is not sufficient
because the nonexistence of the AH is due to the effect of
nonlinear terms in $j$. Thus we should study (at least) the second-order
perturbation, with which we will proceeded in this section.

This study has the following meanings. First, it will clarify to what extent
the condition of AH formation depends on the choice of the slice.
In the previous section, we found that the conditions are very different
in the (2{\it a}) and (2{\it b}) cases. Although we expected that this is
due to the artificial effect of the slice choice, the study in this section
will explicitly show whether such an expectation is correct or not.
Next, by comparing the two cases $\sigma=\pm 1$, 
we can study the properties of the gravitational field
generated by the spin-spin interaction in the gyraton collision.
As a result, we will find
the dependence of the AH formation
on the relative helicities of the incoming gyratons.  
For the old slice $u\ge 0=v$ and $v\ge 0=u$,  
we found no difference between $\sigma=\pm 1$ cases
because the function $G(u,r)$ on the chosen slice depends only on $j^2$. However,
the second-order structure of the region IVA will depend also on $\sigma$
and it will lead to different conditions for the AH formation on the new slice,
which consists of the future boundaries of regions IVA, IIB and IIIB as illustrated
in Fig.~\ref{region-IVA}.

The gravitational spin-spin interactions were studied 
for a spinning test particle around a rotating body \cite{W72},
for a massless particle passing by a rotating body \cite{G02, BG05},
and for binary systems of weakly gravitating bodies \cite{BO75, TH85}.   
In the case of binary systems, the contribution of spin-spin interaction 
to the relative acceleration, $\vec{a}_{\rm SS}$, 
between two bodies
was calculated as
%
\begin{equation}
\vec{a}_{\rm SS}=
-\frac{3}{\mu r^4}
\left[
\vec{n}(\vec{S}_1\cdot\vec{S}_2)
+\vec{S}_1(\vec{n}\cdot\vec{S}_2)
+\vec{S}_2(\vec{n}\cdot\vec{S}_1)
-5\vec{n}(\vec{n}\cdot\vec{S}_1)
(\vec{n}\cdot\vec{S}_2)
\right],
\label{spin-spin-force}
\end{equation}
%
where $\vec{r}$ is a relative location 
$\vec{r}=\vec{r}_1-\vec{r}_2$,
$\vec{n}:=\vec{r}/r$, $\mu$ is the reduced mass, 
$\vec{S}_1$ and $\vec{S}_2$ are spins of two bodies.
For $\vec{S}_1=S_1\vec{n}$ and $\vec{S}_2=S_2\vec{n}$,
the spin-spin acceleration becomes $\vec{a}_{\rm SS}=(6/\mu r^4)S_1S_2\vec{n}$.
Therefore,  
for a binary with both spins aligned 
with the relative location vector $\vec{r}$
(i.e., both $S_1$ and $S_2$ are positive), 
the spin-spin interaction is repulsive.
Another example where the spin-spin
interaction plays an important role is 
the Hawking emission of massless particles with spin
(e.g., photons and neutrinos) by a Kerr black hole. 
In this process, the flux of particles with a given helicity
created by the rotating black hole
has anisotropic distribution.
The black hole radiates more particles in the direction
where the spin is aligned to the angular momentum
of the black hole than in the opposite direction
(\cite{LU79,V79,BF89} and see also 
\cite{IOP03} for higher-dimensional cases).
This indicates the existence of 
a spin-spin interaction between the black hole and
an emitted particle, which is repulsive 
when two spins have the same direction.
If we assume that
a spin-spin interaction similar to the above examples is present
for a system of two relativistic spin particles,
the black hole formation in the head-on collision of
two gyratons with the same spin direction ($\sigma=+1$) is 
expected to be more difficult 
than that with the opposite spin directions ($\sigma=-1$).
The calculations in this section will confirm this.

In Sec. V A, we derive the second-order Einstein equations and solve them.
Then the AH equation and the boundary condition on the new slice is studied
in Sec. V B. We present the numerical results in Sec. V C.
In Sec. V D, we discuss the properties of the gravitational field
in region IVA in more detail using the null geodesics.
This helps us to interpret the results of AH formation.

%
%
\subsection{Second-order equations}

We adopt $\epsilon=2(j/L)$ as a small expansion parameter 
and assume the following metric ansatz in $(u,v,r,\phi)$ coordinates:
%
\begin{equation}
ds^2=-(1+\epsilon^2 c)dudv+(1+\epsilon^2 a)dr^2
+r^2(1+\epsilon^2 b)d\phi^2+2\epsilon\frac{u+\sigma v}{r}drd\phi.
\label{ansatz}
\end{equation}
%
Here $a,b$ and $c$ are functions of $u,v$ and $r$.
Expanding the Einstein equations up to the second order in $\epsilon$,
we obtain\footnote{
More strictly, we should put 
$g_{r\phi}=\epsilon [u\theta(u)+\sigma v\theta(v)]/r$
in Eq.~\eqref{ansatz}, although we do not show this
because the equations become
tedious. We note that this step function formula 
leads to the junction conditions at 
$u=0\le v\le L$ and $v=0\le u\le L$ through
$R_{uu}=R_{vv}=0$, which
the solution \eqref{solution-a}--\eqref{solution-c} satisfies.}:
%
\begin{align}
a_{,uu}+b_{,uu}&=
\frac{1}{r^4},
\label{uu}
\\
a_{,vv}+b_{,vv}&=\frac{\sigma^2}{r^4},
\label{vv}
\end{align}
%
%
\begin{equation}
2c_{,uv}-\frac12\left(c_{,rr}+\frac{c_{,r}}{r}\right)
+a_{,uv}+b_{,uv}=\frac{\sigma}{r^4},
\label{uv}
\end{equation}
%
%
\begin{align}
c_{,ur}+b_{,ur}+\frac{b_{,u}}{r}&=\frac{a_{,u}}{r}
-\frac{2}{r^5}(u+\sigma v),
\label{ur}
\\
c_{,vr}+b_{,vr}+\frac{b_{,v}}{r}&=\frac{a_{,v}}{r}
-\frac{2\sigma}{r^5}(u+\sigma v),
\label{vr}
\end{align}
%
%
\begin{align}
2c_{,rr}-4a_{,uv}+b_{,rr}+\frac{2}{r}b_{,r}&=\frac{a_{,r}}{r}
+\frac{4}{r^6}(u+\sigma v)^2-\frac{4\sigma}{r^4},
\label{rr}
\\
2\frac{c_{,r}}{r}-4b_{,uv}+b_{,rr}+\frac{2}{r}b_{,r}&=\frac{a_{,r}}{r}
+\frac{4}{r^6}(u+\sigma v)^2-\frac{4\sigma}{r^4}.
\label{phiphi}
\end{align}
%
These relations follow from $uu,~vv,~uv,~ur,~vr,~rr,~\phi\phi$ components
of the equation $R_{\mu\nu}=0$, respectively. 
The other components vanish
automatically.

The initial conditions for this system
 are found by expanding the exact metric
in regions IIA and IIIA in terms of $j$:
%
\begin{align}
a=b=\frac{u^2}{4r^4},~~c=0,~~&\textrm{for}~~v=0,
\\
a=b=\frac{\sigma^2v^2}{4r^4},~~c=0,~~&\textrm{for}~~u=0.
\end{align}
%
The solutions satisfying these initial conditions are found as
%
\begin{equation}
a=\frac{1}{4r^4}(u+\sigma v)^2
+\frac{\sigma}{2r^2}\left[x\left(3-\frac{1}{1-x}\right)+\log(1-x)\right],
\label{solution-a}
\end{equation}
%
%
\begin{equation}
b=\frac{1}{4r^4}(u+\sigma v)^2
-\frac{\sigma}{2r^2}\left[x\left(1-\frac{1}{1-x}\right)+\log(1-x)\right],
\label{solution-b}
\end{equation}
%
%
\begin{equation}
c=-\frac{\sigma}{2r^2}\frac{x}{1-x},
\label{solution-c}
\end{equation}
%
with
%
\begin{equation}
x=uv/r^2.
\end{equation}
%

We discuss now the properties of the second-order solution
in region IVA.
First, a line $u,r,\phi=\mathrm{const}$ is a null geodesic, 
although when $c\neq 0$
the coordinate $v$ is no longer an affine parameter 
along the geodesic.
Similarly a line $v,r,\phi=\mathrm{const.}$ is a null geodesic, 
although the coordinate $u$ is not an affine parameter along it. 
Thus, the coordinates $(u,v,r,\phi)$ 
simultaneously label the two null geodesic congruences. 

Next, all second-order quantities 
$a$, $b$, and $c$ diverge at $x=1$, i.e., $uv=r^2$. 
Therefore it is interesting to ask whether $x=1$ is a physical singularity
or a coordinate singularity.
For this purpose, we calculated
the leading term in the expansion of the Kretchman invariant
$K:= R_{abcd}R^{abcd}$ near this point:
%
\begin{equation}
K=
\epsilon^4\sigma^2\frac{4(3-x)}{r^8(1-x)^3}.
\label{Riemann-invariant}
\end{equation}
%
Evidently it is divergent at $x=1$. 
Because we are studying perturbation,  
the formula \eqref{Riemann-invariant} cannot be
trusted in the neighborhood of $x=1$, and we cannot
definitely claim that there is a physical singularity at $x=1$.
Still, Eq.~\eqref{Riemann-invariant}
indicates that there always exists the region
where the perturbation breaks down around $x=1$
for any small $j$. Hence,
it is natural to expect that 
the exact solution, if it is found, also will have 
a real singularity of which location is shifted by $O(j^2)$ from $x=1$.
If this is the case, a physical 
singularity is produced at $u=v=r=0$
by the collision of gyratons and expands (almost) at
the speed of light 
because $uv=r^2$ represents a light cone in the background spacetime.
We note that this singularity formation is a consequence of the infinitely narrow shape 
of the source, i.e. Eqs.~\eqref{Tuu} and \eqref{Tua}. 
In a realistic case where 
each source has a finite radius $\bar{r}=\bar{r}_{\rm s}$,
the metric is regular at the source and
therefore the singularity is not produced at $uv=r^2$. 
Then the spacetime structure in the region $uv>r^2$
is determined by matter interaction between the two sources.
Although the dependence on the properties of matter interaction 
is an interesting issue, it is not tractable
by the perturbation.

Finally, although the metric is continuous
everywhere, first derivatives of $a$, $b$, and $c$ are discontinuous 
at $u=0\le v\le L$ and $v=0\le u\le L$.  
As a result, some components of Riemann curvature, $R_{urur}$,
$R_{vrvr}$, $R_{u\phi u\phi}$ and $R_{v\phi v\phi}$, have
the delta function singularity there:
%
\begin{eqnarray}
R_{urur}&=&\epsilon^2\frac{\sigma}{4r^4}
\left[
\frac{v^2(3-x)}{r^2(1-x)}\theta(u)
-{2v}\delta(u)
\right]\theta(v),
\label{Riemann-urur}
\\
R_{u\phi u\phi}&=&\epsilon^2\frac{\sigma}{4r^4}
\left[
\frac{v^2(3-x)}{r^2(1-x)}\theta(u)
+{2v}\delta(u)
\right]\theta(v),
\label{Riemann-uphiuphi}
\end{eqnarray}
%
and $R_{vrvr}$ and $R_{v\phi v\phi}$ are obtained by changing 
$u$ and $v$ in Eqs.~\eqref{Riemann-urur}
and \eqref{Riemann-uphiuphi}, respectively
(but note that Ricci tensor is zero in the sense of distribution).
Hence, at the encounter of the two spin flows,
a new shock field is produced and it grows linearly in $u$ or $v$.
The above four components of Riemann curvature
are proportional to $\sigma$ and thus the feature of the shock gravitational field
at $u=0\le v\le L$ and $v=0\le u\le L$ depends on the sign of $\sigma$.

%
%
\subsection{AH equation on the new slice}

Because the metric in region IVA has properties
that are somewhat different from other regions,
we should derive the AH equation on the new slice.
But the basic idea is the same as that in Sec. II.

%
\begin{figure}[tb]
 \centering 
  \includegraphics[width=0.5\textwidth]{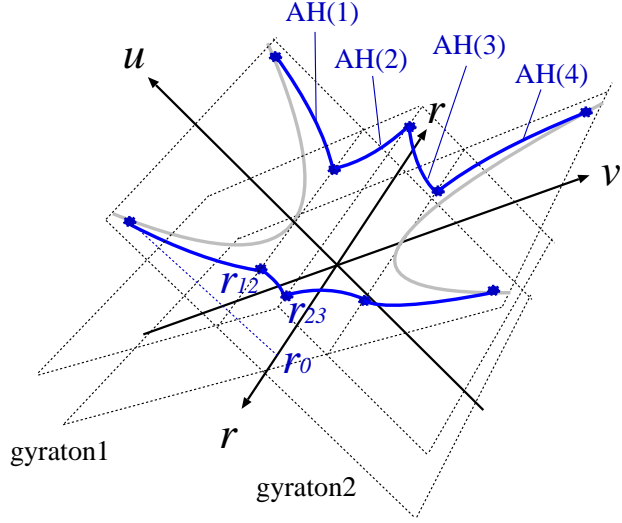}
 \caption{The new slice on which we study the AH formation and
 the schematic shape of AH in the new slice.}
 \label{new-slice}
\end{figure}
%

The second-order metric can be written like
%
\begin{equation}
ds^2=-Cdudv+Adr^2+B(d\phi+D dr)^2,
\end{equation}
%
where $A,B,C$ and $D$ are functions of $u,v$ and $r$.
In region IVA, 
%
\begin{equation}
A=1+\epsilon^2\left[a-\frac{(u+\sigma v)^2}{r^4}\right],~
B=r^2(1+\epsilon^2 b),~
C=1+\epsilon^2 c,~
D=\epsilon\frac{(u+\sigma v)}{r^3},
\end{equation}
%
and in region IIB,
%
\begin{eqnarray}
A&=&\left(1+\frac{u_L}{r^2}\right)
\left\{
1+\frac{u_L}{r^2}
-\frac{j^2}{r^2}
\left[
3
+\left(\frac{5}{r^2}+\frac{6}{L}\right)u_L
\right]
\right\},
\nonumber\\
B&=&
\left(r-\frac{u_L}{r}\right)
\left\{
r-\frac{u_L}{r}
+\frac{j^2}{r^3}
\left[
1+\left(\frac{1}{r^2}+\frac{2}{L}\right)u_L
\right]
\right\},
\nonumber\\
C&=&1, \ D\ =\  \frac{2j}{r^3},
\end{eqnarray}
%
where $u_L:=u-L$ and we have kept terms up to second order in $j$.
Based on this metric, we solve the AH equation on the new slice
as shown in Fig. \ref{new-slice}.
The new slice consists of four parts: 
(1) $v=0,\ L\le u$; 
(2) $u=L,\ 0\le v\le L$; 
(3) $v=L,\ 0\le u\le L$;
(4) $u=0,\ L\le v$. 
Correspondingly, the AH consists of 
$u=h^{(1)}(r)$ on the slice (1); 
$v=h^{(2)}(r)$ on the slice (2); 
$u=h^{(3)}(r)$ on the slice (3); 
$v=h^{(4)}(r)$ on the slice (4).

We consider the AH equation for $h(r):=h^{(2)}(r)$.
The tangent vector of the null geodesic congruence
from the surface is given by
%
\begin{equation}
\left(k^u, k^v, k^r, k^\phi\right) 
=
\left( 
2, \frac{C}{2A}h_{,r}^{2}, \frac{C}{A}h_{,r}, -\frac{CD}{A}h_{,r}
\right).
\end{equation} 
%
and the condition of zero expansion 
becomes 
%
\begin{equation}
\partial_rk^r+\frac{A_{,u}k^u+A_{,v}k^v+A_{,r}k^r}{2A}
+\frac{B_{,u}k^u+B_{,v}k^v+B_{,r}k^r}{2B}=0,
\end{equation}
%
or equivalently
%
\begin{equation}
h_{,rr}
+\left(-3\frac{A_{,v}}{A}+\frac{B_{,v}}{B}+4\frac{C_{,v}}{C}\right)
\frac{h_{,r}^{2}}{4}
+\left(-\frac{A_{,r}}{A}+\frac{B_{,r}}{B}+2\frac{C_{,r}}{C}\right)
\frac{h_{,r}}{2}
+\frac{A}{C}
\left(\frac{A_{,u}}{A}+\frac{B_{,u}}{B}\right)
=0.
\label{AH-eq.}
\end{equation}
%
The equation for $h^{(1)}(r)$ is obtained by just changing $u$ and $v$ in Eq. \eqref{AH-eq.}.

Now we explain the boundary conditions.
At the intersection of the AH and the coordinate 
singularity $u=L+r^2$, we impose
%
\begin{equation}
h^{(1)}_{,r}=-2{B_{,r}}/{B_{,u}}.
\label{boundary0}
\end{equation}
%
with the same reason as that in Sec. II. 
At the intersection of slices (1) and (2), we impose the condition
that two null vectors of both sides of the surface be parallel, which is equivalent to
%
\begin{equation}
h^{(1)}_{,r}h^{(2)}_{,r}=4A,
\label{boundary12}
\end{equation}
%
where we used $C=1$ on $v=0$.
Similarly, at the intersection of slices (2) and (3), we impose
%
\begin{equation}
h^{(2)}_{,r}h^{(3)}_{,r}={4A}/{C}.
\label{boundary23-pre}
\end{equation}
%

In the cases $\sigma=\pm 1$, 
the functions $A, B$ and $C$ are symmetric with respect to $u$ and $v$.
Because $D$ does not appear in the AH equation and the boundary conditions,
the AH shape is symmetric with respect to the plane $u=v$. 
Hence, we only have to study
$h^{(1)}(r)$ and $h^{(2)}(r)$, and the boundary condition
\eqref{boundary23-pre} is reduced to 
%
\begin{equation}
h^{(2)}_{,r}=2\sqrt{{A}/{C}}.
\label{boundary23}
\end{equation} 
%
We also note that because 
the functions $A$, $B$, and $C$
do not depend on the sign of $j$, 
the condition for the AH formation
is written in terms of $|j|$ and $L$
for each $\sigma$. 
For this reason, $j$ is assumed to be positive
without loss of generality in the following.

The numerical procedure is as follows. First we choose some value
of $r_0$ and start solving $h^{(1)}(r)$ with the boundary condition
$h^{(1)}(r_0)=L+r_0^2$ and
\eqref{boundary0}. 
When $h^{(1)}(r)$ becomes $L$ at $r=r_{12}$, we solve $h^{(2)}$
using the boundary conditions $h^{(2)}(r_{12})=0$ and $h^{(2)}_{,r}=4A/h^{(1)}_{,r}$.
When $h^{(2)}$ becomes $L$ at $r=r_{23}$, we check whether the 
boundary condition \eqref{boundary23} is satisfied.
Iterating these steps for various values of $r_0$, we can judge the
existence of the AH and find its location.

%
%
\subsection{Numerical results}

%
\begin{figure}[tb]
 \centering 
 \includegraphics[width=0.3\textwidth]{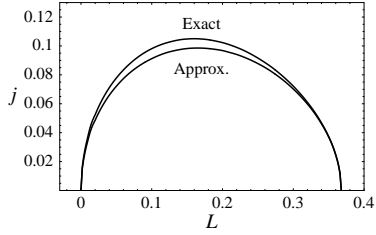}
 \caption{The critical lines for AH formation on the $(L, j)$-plane
 for the old slice.
 The results obtained by the exact formula  and by the second-order
 approximation (denoted by ``Exact'' and ``Approx.'', respectively)
 are compared. The two results agree well and the error is $O(j^2).$}
 \label{compare-2O-exact}
\end{figure}
%

In order to test the reliability of the second-order approximation,
we studied the condition of AH existence on the
old slice $u\ge 0=v$ and $v\ge 0=u$
using the exact formula and the second-order formula for $G(u,r)$
and compared the two results. 
Figure \ref{compare-2O-exact} shows the regions of AH formation 
on the $(L, j)$-plane
obtained by the two methods.
The two results agree well and the difference
is $O(j^2)$. This demonstrates the reliability
of the second-order approximation for the old slice.
Later, we will discuss also the reliability of the approximation
for the AH study on the new slice.

%
\begin{figure}[tb]
 \centering 
 \includegraphics[width=0.2\textwidth]{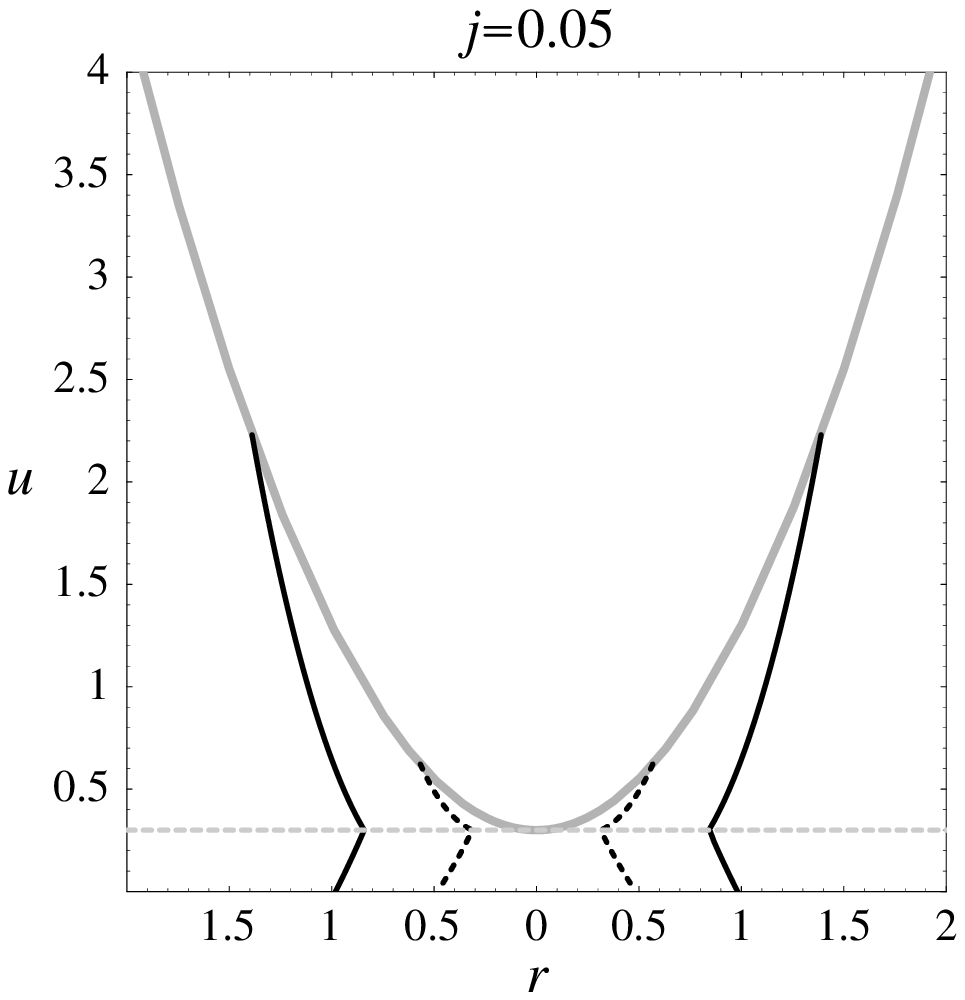}
 \includegraphics[width=0.2\textwidth]{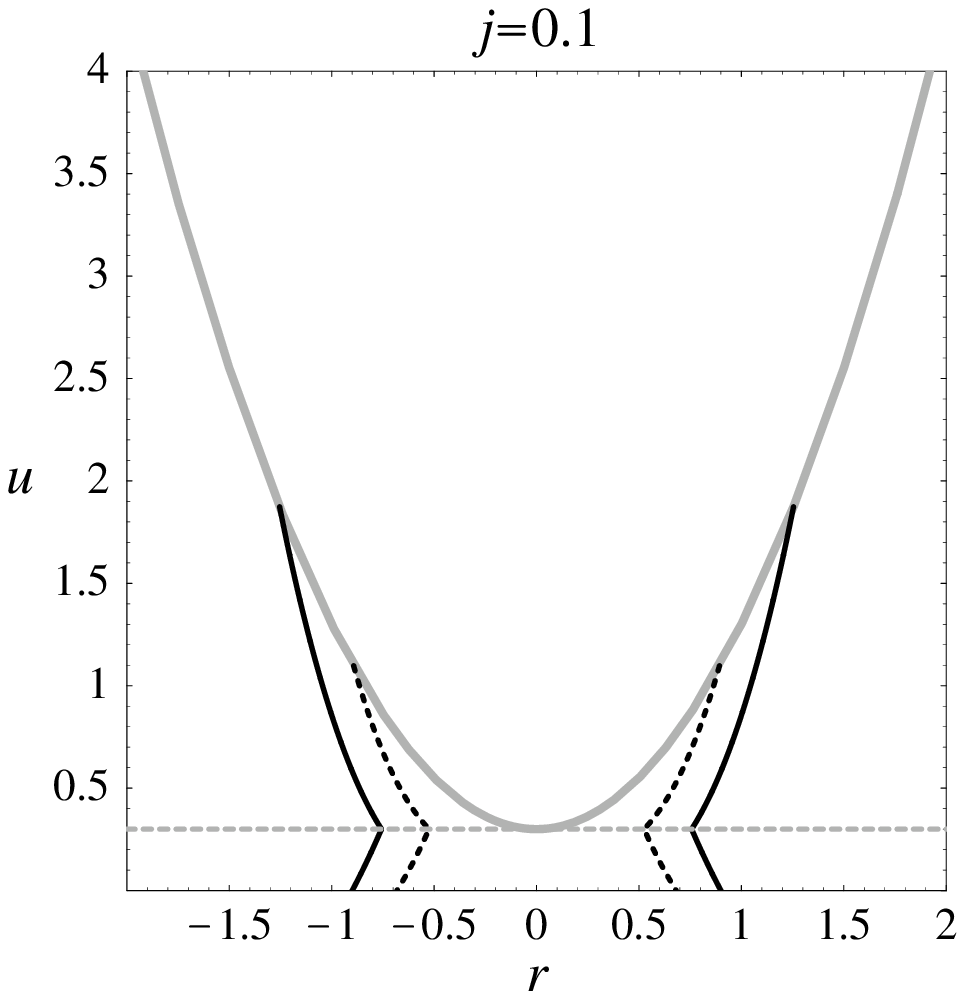}
 \includegraphics[width=0.2\textwidth]{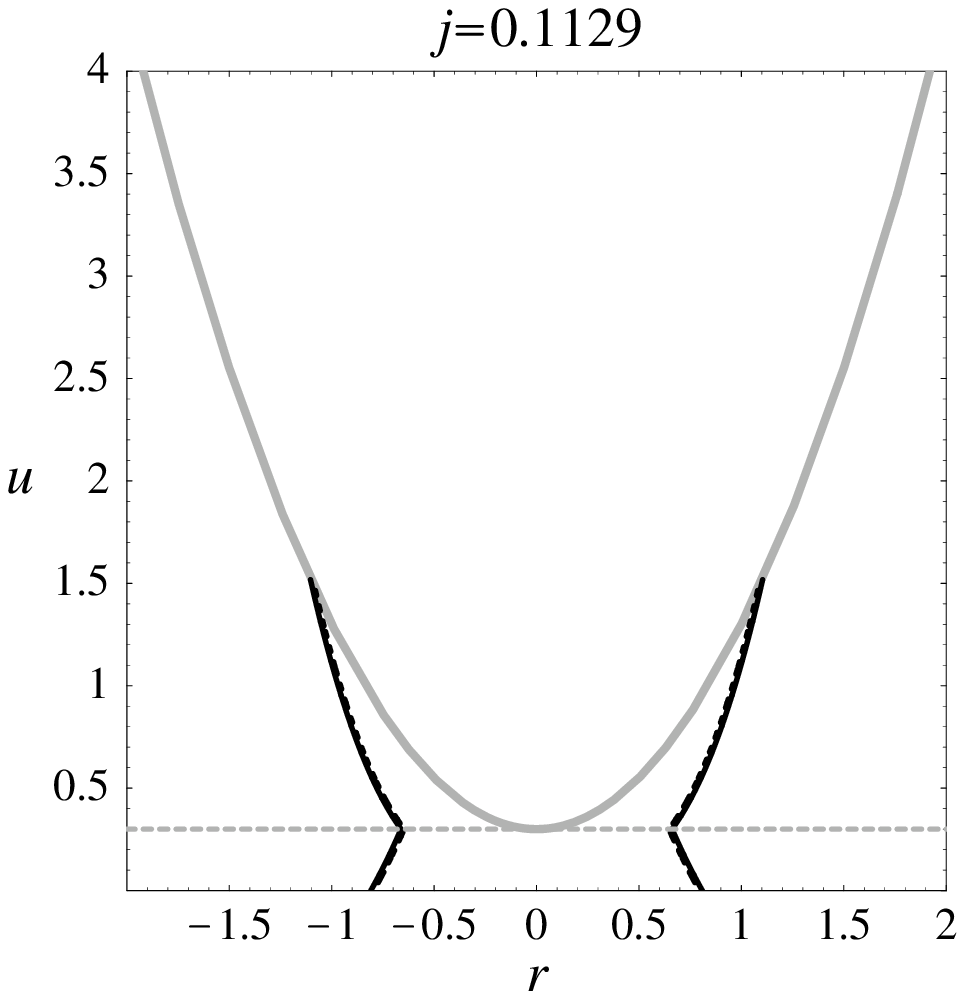}
 \caption{The top view of the AH (solid lines) and the inner boundary of 
 the trapped region
 (dashed lines) in the case $\sigma=+1$ for $L=0.3$ and $j=0.05, 0.1$, and $0.1129$.}
 \label{AH-2O-jj}
\end{figure}
%

Now we show the results of the new slice. 
We first show the case $\sigma=+1$
where the spins of two gyratons
have the same direction 
(i.e., the helicity of one gyraton is positive and that of the other is negative).
Figure \ref{AH-2O-jj} shows top views of the AH shape for
$L=0.3$ and $j=0.05, 0.1$ and $0.1129$. 
We could not find the solution for $j\ge 0.1130$.
Similarly to the case of the old slice, there is 
some critical value of the spin $j_{\rm crit}^{(+)}(L)$ for the AH formation.
There are two solutions for each $j<j_{\rm crit}^{(+)}(L)$,
which correspond to the AH and the inner
boundary of the trapped region, 
and they coincide at $j=j_{\rm crit}^{(+)}(L)$.

%
\begin{figure}[tb]
 \centering 
 \includegraphics[width=0.2\textwidth]{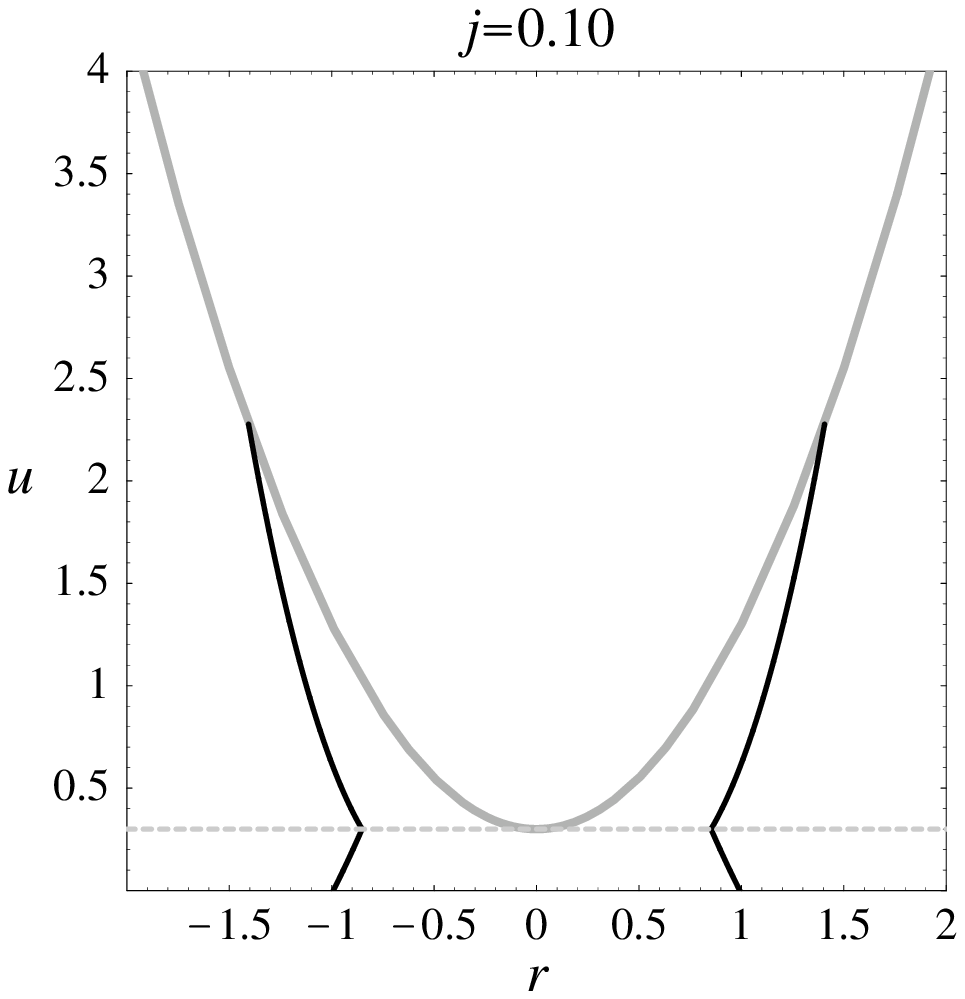}
 \includegraphics[width=0.2\textwidth]{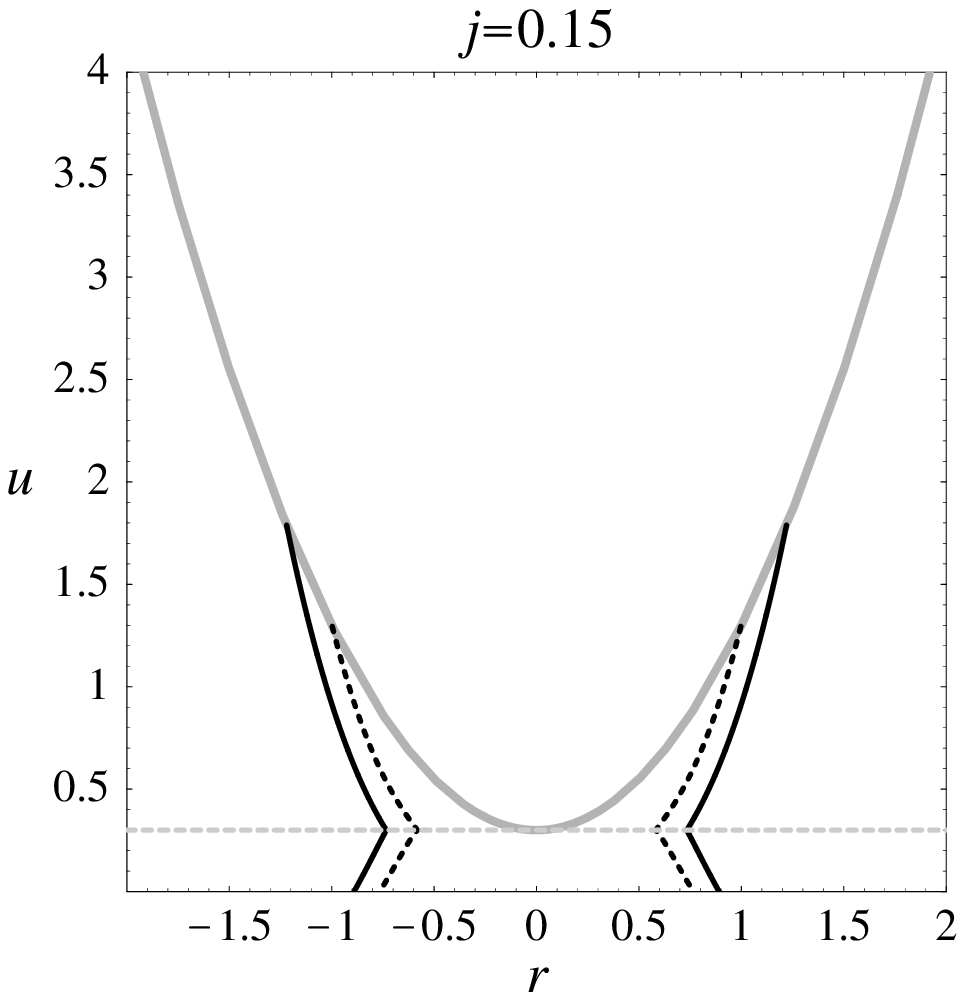}
 \includegraphics[width=0.2\textwidth]{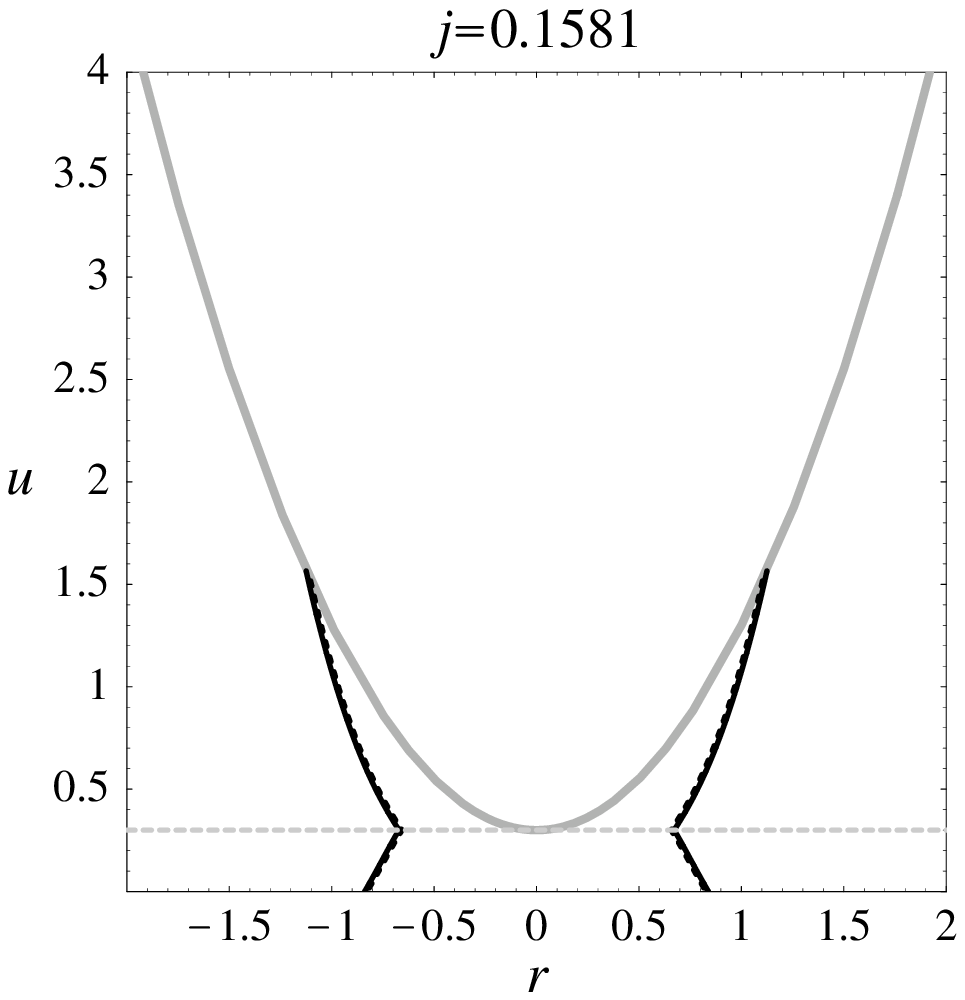}
 \caption{The top view of the AH (solid lines) and the inner boundary of trapped region
 (dashed lines) in the case $\sigma=-1$ for $L=0.3$ and $j=0.1, 0.15$, and $0.1581$.
 For $j=0.1$, there is no inner boundary of trapped region.}
 \label{AH-2O-jmj}
\end{figure}
%

Next we show the case $\sigma=-1$,
where the spins of two gyratons are oppositely directed
(i.e., the helicities of gyratons are both positive or negative).
Figure~\ref{AH-2O-jmj} shows top-views of the 
AH shape for $L=0.3$ and $j=0.1, 0.15, 0.1581$. 
For $j\lesssim 0.1$, only one solution is found.
Hence there is an AH but 
no inner boundary of the trapped region. 
For $0.15\lesssim j\le 0.1581$, 
two solutions are found for each $j$. Thus the inner
boundary of the trapped region appears for these values of $j$. 
For $j\ge 0.1582$, there was no solution. 
The critical value $j_{\rm crit}^{(-)}(L)$ of AH formation
for $\sigma=-1$ is larger than $j_{\rm crit}^{(+)}(L)$. 
Thus in the case $\sigma=-1$, the AH is allowed 
to form in a larger parameter regime compared to the case $\sigma=+1$.

We studied the critical value $j_{\rm crit}^{(\sigma)}(L)$ as 
functions of $L$ for the cases $\sigma=\pm 1$. 
Before showing the obtained results, 
we comment on the reliability of the second-order approximation.
In order to evaluate the error, we checked
the maximum values of $\epsilon^2 a, \epsilon^2b$ and $ \epsilon^2c$
on the AH at the critical line. 
In the case $\sigma=+1$, 
$\mathrm{max}[\epsilon^2a]\le 0.32$, 
$\mathrm{max}[\epsilon^2b]\le 0.20$, and 
$\mathrm{max}[\epsilon^2c]\le 0.14$
are satisfied for arbitrary $L$. 
Therefore the expected error is about $20\%$.
In the case $\sigma=-1$,
they are found to be 
$\mathrm{max}[\epsilon^2a]\simeq 0.38$,
$\mathrm{max}[\epsilon^2b]\simeq 0.13$ and 
$\mathrm{max}[\epsilon^2c]\simeq 0.12$
for $L\lesssim 0.3$, and thus the expected error 
is about $30\%$ for $L\lesssim 0.3$. 
However, for $L=1.0$,
their values grow large:
$\mathrm{max}[\epsilon^2a]\simeq  0.96$, 
$\mathrm{max}[\epsilon^2b]\simeq 0.97$ and 
$\mathrm{max}[\epsilon^2c]\simeq 0.95$,
and the approximation obviously breaks down at $L\gtrsim 0.9$.
Thus unfortunately we cannot trust the shape of the critical line 
for $L\gtrsim 0.9$ in the case $\sigma=-1$.
To summarize, we can trust the shape of the AH critical line
of $\sigma=+1$ for arbitrary $L$ and that of $\sigma=-1$ for $L\lesssim 0.3$
with the error discussed above.

%
\begin{figure}[tb]
 \centering 
 \includegraphics[width=0.4\textwidth]{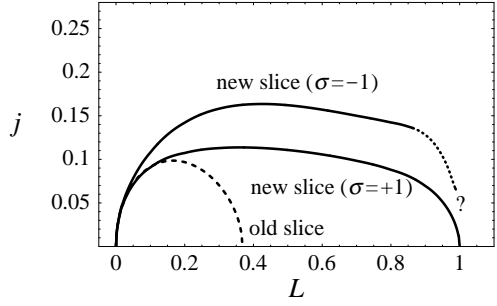}
 \caption{The critical lines for the AH formation for the new slice for $\sigma=+1$
 (lower solid line) and for $\sigma=-1$ (upper solid line). The critical line for the old
 slice is also shown (dashed line). In both cases $\sigma=\pm 1$, the AH formation
 is allowed in a larger region on the $(L,j)$-plane compared to the old slice.
 The allowed region of $\sigma=-1$ is larger than that of $\sigma=+1$. 
 In the case $\sigma=-1$, the perturbative quantity becomes large
 at $L\gtrsim 0.9$ and the shape of the critical line cannot be trusted there (dotted line).
 }
 \label{secondorder-jmj}
\end{figure}
%

Figure~\ref{secondorder-jmj} shows the parameter regions in 
the $(L,j)$-plane
that allows the AH formation on the new slice in the cases 
$\sigma=\pm 1$, together with that on the old slice.
In both cases, $j_{\rm crit}^{(\sigma)}(L)$ goes to zero 
in the limit $L\to 0$. 
The critical line of $\sigma=+1$ crosses the $L$-axis at $L=1$, 
which is much larger compared to $L=1/e$ in the case of the old slice.
Although the error in $j_{\rm crit}^{(-)}(L)$ grows large
for $L\gtrsim 0.9$, the critical line of $\sigma=-1$ does not 
seem to cross the $L$-axis for $0\le L< 1$. Hence,  
the allowed regions of the new slice is 
much larger than that of the old slice for both $\sigma=\pm 1$. 
At the end of the previous section, we stated our expectation
that the large difference between the allowed regions of the (2{\it a}) and (2{\it b})
cases is due to the artificial effect of the slice choice.
It is now confirmed, since the allowed region of 
the (2{\it b}) case has become much larger by just changing the slice.
Comparing the two cases $\sigma=\pm 1$,
$j_{\rm crit}^{(-)}(L)$ is greater than $j_{\rm crit}^{(+)}(L)$. 
Therefore, the AH formation in the case $\sigma=-1$ is allowed 
in a larger parameter region compared to the case $\sigma=+1$,
and the condition of the AH formation depends on the relative helicities of
incoming gyratons. 
To briefly summarize, on the new slice,
the condition of the AH formation is roughly expressed as 
$L\sim 0.5$ and $j\lesssim 0.1$ in the case $\sigma=+1$
and $L\sim 0.5$ and $j\lesssim 0.15$ in the case $\sigma=-1$.

The reason why the allowed region in the case $\sigma=+1$ 
is limited to $0\le L\le 1$ is understood as follows.
In the case $j=0$, the AH solution is given by
%
\begin{equation}
h^{(1)}(r)=h^{(4)}(r)=L+2r^2\log\left(r/r_{12}\right),~~(r_{12}\le r\le \sqrt{e}r_{12}),
\label{solution-2O-jeq0-1}
\end{equation}
%
%
\begin{equation}
h^{(2)}(r)=h^{(3)}(r)=2\log\left(r/r_{12}\right),~~(r_{12}\le r\le 1),
\label{solution-2O-jeq0-2}
\end{equation}
%
with $r_{12}=1/\sqrt{L}$. Although the AH is expected 
to converge to this solution in the limit $j\to 0$, 
we should take care of the presence
of the singularity $uv=r^2$, where the perturbative quantities diverge. 
For $L>1$, the singularity 
crosses the surface \eqref{solution-2O-jeq0-2}, 
invalidating it to be an AH. Hence $j=0$ and $j=0_+$ are different
for $L>1$, and no AH exists for small $j$.
On the other hand, for $0\le L\le 1$,
the surface \eqref{solution-2O-jeq0-1} and \eqref{solution-2O-jeq0-2} 
is the AH in the limit $j\to 0$, because the singularity does not
cross the surface. Hence, it is natural that
the region of the AH formation is restricted to $0\le L\le 1$.
Although we could not specify the allowed region for $\sigma=-1$ 
around $L\simeq 1$,
the above discussion would hold also for this case. Hence, 
if the exact solution for region IVA is found, 
the allowed region for $\sigma=-1$ will turn out to be restricted to $0\le L\le 1$
\footnote{This discussion holds only for a
collision of gyratons with singular sources, Eqs.~\eqref{Tuu} and \eqref{Tua}.
In the collision of realistic beam pulses, 
the singularity is not produced at $uv=r^2$ and
the regions of AH formation 
might become different from Fig.~\ref{secondorder-jmj}.
}.

%
%
\subsection{Gravitational field in the region IVA}

We discuss the properties of gravitational field 
in region IVA in more detail,
because it helps us to understand the reason for the different allowed
regions in the cases $\sigma=\pm 1$.
For this purpose, we study the ``gravitational force'' acting on
the null geodesics
$u,r,\phi=\mathrm{const.}$ and $v,r,\phi=\mathrm{const.}$

Let us consider a null geodesic congruence $u=u_0$, $r=r_0$, $0\le \phi\le 1$.
The section of the congruence and $v=\mathrm{const.}$ is a loop
and the quantity
%
\begin{equation}
r_{\rm loop}^{(u_0,r_0)}(v)=r_0(1+\epsilon^2b(u_0, v, r_0)/2)
\end{equation}
%
gives a radius of the loop (i.e., the proper circumference
divided by $2\pi$). We define the ``gravitational force'' 
$F^{(u_0,r_0)}(v)$ toward the symmetry axis by 
%
\begin{equation}
F^{(u_0,r_0)}(v):=\frac{\partial^2 r_{\rm loop}^{(u_0,r_0)}}{\partial v^2}.
\end{equation}
%
The force is attractive if $F^{(u_0,r_0)}(v)<0$ and 
repulsive if $F^{(u_0,r_0)}(v)>0$. Similarly we consider 
another congruence
$v=v_0$, $r=r_0$, $0\le \phi\le 1$ 
and introduce its loop radius $r_{\rm loop}^{(v_0,r_0)}(u)$.
Then another kind of force is defined by 
%
\begin{equation}
F^{(v_0,r_0)}(u):=\frac{\partial^2 r_{\rm loop}^{(v_0,r_0)}}{\partial u^2}.
\end{equation}
%
The two forces are calculated as
%
\begin{equation}
F^{(u,r)}(v)=
\frac{\epsilon^2}{4r^3}
\left[
2\sigma{u}\delta(v)
+
1+\sigma\frac{u^2(3-x)}{r^2(1-x)^3}
\right],
\label{Force-ur-v}
\end{equation}
%
%
\begin{equation}
F^{(v,r)}(u)=
\frac{\epsilon^2}{4r^3}
\left[
2\sigma{v}\delta(u)
+
1+\sigma\frac{v^2(3-x)}{r^2(1-x)^3}
\right].
\label{Force-vr-u}
\end{equation}
%
The delta function of the first term in the square brackets of each formula
comes from the new shock field
at $u=0\le v\le L$ and $v=0\le u\le L$ 
[see Eqs. \eqref{Riemann-urur} and \eqref{Riemann-uphiuphi}].

In the case $\sigma=+1$,
both $F^{(u,r)}(v)$ and $F^{(v,r)}(u)$
are positive outside of the singularity $x=1$.
Hence, the gravitational field is repulsive in the whole region IVA.
On the other hand, in the case $\sigma=-1$, the coefficients of
the delta functions in Eqs.~\eqref{Force-ur-v} and \eqref{Force-vr-u}
are negative, indicating that the new shock fields are attractive.
The third term in the square brackets is also negative. If $x$ is close enough to 1, 
the third term exceeds the second term
and the force becomes negative. Hence, around the singularity $x=1$,
there is always the attractive region of the gravitational force.
If $x$ is close to $0$, the third term 
is smaller than 1 and 
the gravitational field is repulsive in such a region.  
Therefore, both attractive and repulsive regions 
exist for $\sigma=-1$.

%
\begin{figure}[tb]
 \centering 
 \includegraphics[width=0.4\textwidth]{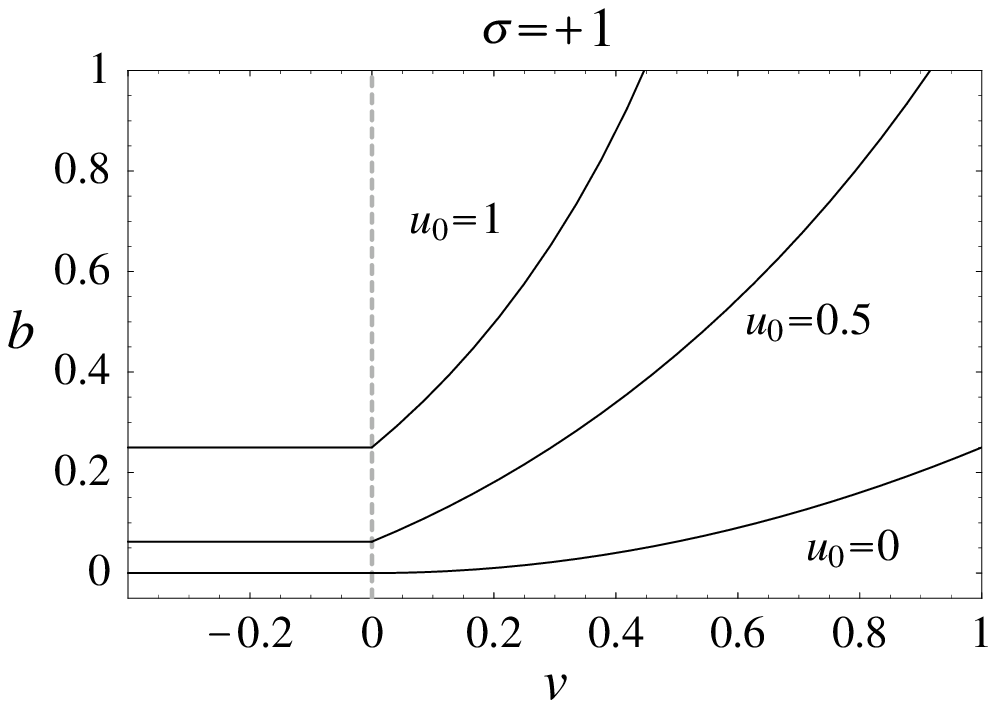}
 \hspace{10mm}
 \includegraphics[width=0.4\textwidth]{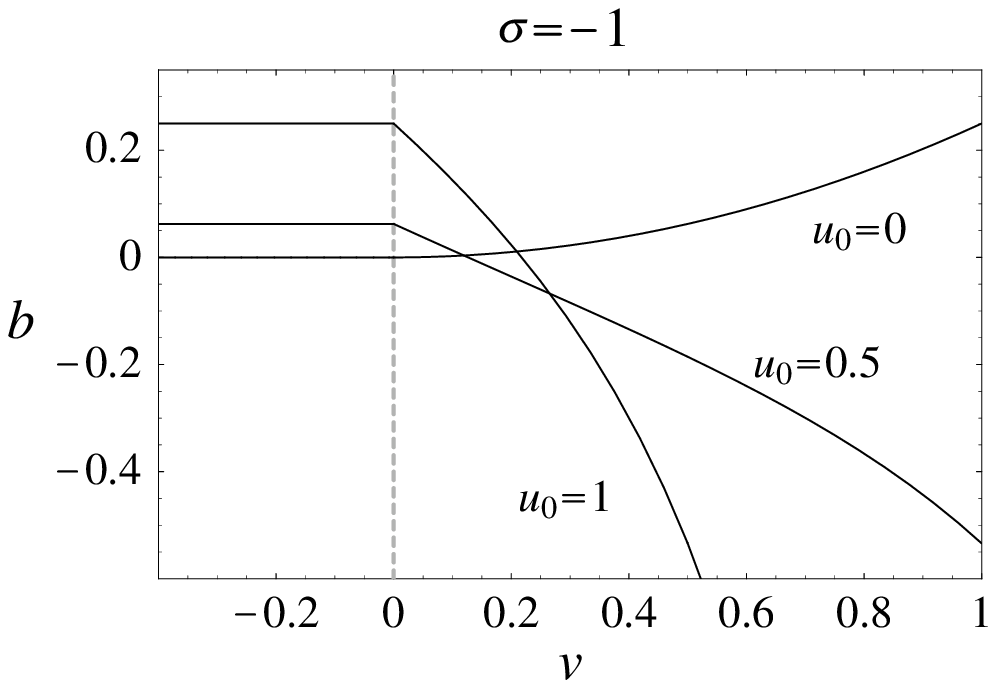}
 \caption{
 The behavior of the function $b(u_0, v, r_0)$
 in the cases $\sigma=+1$ (left) and $\sigma=-1$ (right)
 for $r_0=1$ and $u_0=0, 0.5, 1$. The value of
 $b(u_0, v, r_0)$ is directly related to the radius $r_{\rm loop}^{(u_0, r_0)}(v)$ 
 of the light ray congruence 
 $u=u_0$, $r=r_0$, $0\le \phi\le 2\pi$.
 The light rays quickly bend at $v=0$ 
 due to the delta functions in Riemann curvature 
 \eqref{Riemann-urur} and \eqref{Riemann-uphiuphi} in both cases 
 but the bending directions are opposite. 
 $b(u_0, v, r_0)$ continues to increase in the case $\sigma=+1$,
 while its behavior strongly depends on $u_0$ in the case $\sigma=-1$. 
 }
 \label{b-evolution}
\end{figure}
%

Let us look at the behavior of the loop radius $r_{\rm loop}^{(u_0, r_0)}(v)$.
Ignoring a factor, the change in $r_{\rm loop}^{(u_0, r_0)}(v)$
is presented by $b(u_0, v, r_0)$. 
Figure \ref{b-evolution} shows the behavior of $b(u_0,v,r_0)$
for $r_0 = 1$ and $u_0 = 0, 0.5, 1$ for the two cases $\sigma=\pm 1$. 
Because of the delta function in the force~\eqref{Force-ur-v},  
$b(u_0, v, r_0)$ is not smooth 
at $v = 0$ for $u_0 > 0$ in both cases.
In the case $\sigma=+1$, $b(u_0,v,r_0)$ suddenly increases at $v=0$
and blows up, since the force is repulsive everywhere.
In the case $\sigma = -1$,
$b(u_0,v,r_0)$ suddenly decreases at $v=0$
due to the attractive force.
For $v>0$,  the behavior of
$b(u_0,v,r_0)$ strongly depends on the value of $u_0$.
If $u_0$ is large, the light ray feels repulsive force at the beginning
but later feels attractive force.

%
\begin{figure}[tb]
 \centering 
 \includegraphics[width=0.4\textwidth]{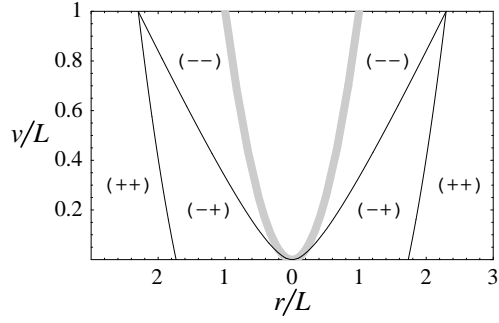}
 \caption{
The sign of two forces 
$F^{(u,r)}(v)$ and $F^{(v,r)}(u)$ on slice (2), i.e. $0\le v\le L=u$,
for $\sigma=-1$.
The slice is divided into three regions: 
region $(--)$ where $F^{(u,r)}(v)<F^{(v,r)}(u)<0$, 
region $(-+)$ where $F^{(u,r)}(v)<0<F^{(v,r)}(u)$,
and region $(++)$ where $0<F^{(u,r)}(v)<F^{(v,r)}(u)$. 
The unit of $r$ (horizontal line) and $v$ (vertical line) is the spin duration $L$
and the gray line indicates the singularity $x=1$.
 }
 \label{forcefield-slice2}
\end{figure}
%

Now we discuss the reason for $j_{\rm crit}^{(-)}(L) \ge  j_{\rm crit}^{(+)}(L)$,
i.e., the difference between the allowed regions 
for the AH formation of $\sigma=\pm1$.
In the case $\sigma = +1$, the gravitational field in the region IVA is
repulsive everywhere. If $j$ is increased, the repulsive force
exceeds the attractive force generated by the energy, 
causing the disappearance of the trapped region.
On the other hand, in the case $\sigma = -1$ 
there are both attractive and repulsive regions.
Figure~\ref{forcefield-slice2} shows the sign of two forces 
$F^{(u,r)}(v)$ and $F^{(v,r)}(u)$ on slice (2), i.e. $0\le v\le L=u$.
The slice is divided into three regions: 
region $(--)$ where $F^{(u,r)}(v)<F^{(v,r)}(u)<0$, 
region $(-+)$ where $F^{(u,r)}(v)<0<F^{(v,r)}(u)$,
and region $(++)$ where $0<F^{(u,r)}(v)<F^{(v,r)}(u)$. 
This figure shows that the gravitational force is attractive 
around the singularity $x=1$ and repulsive for $r\gg L$.
For $L\ll 1$, the attractive region
is a tiny portion just around the singularity
and the force is repulsive almost everywhere on the surface.
The repulsive force becomes dominant as $j$ is increased,
resulting in disappearance of the AH.
Although the attractive region becomes large for $L\simeq 1$,
the attractive force does not help the AH formation effectively
since there is the constraint $j_{\rm crit}^{(-)}(L=1)=0$
coming from the size of the singularity $x=1$
as discussed in the previous subsection.
Therefore, also in the case $\sigma = -1$,
the spin $j$ makes the AH formation more difficult.
However, in the case $\sigma=-1$, 
the repulsive force is obviously smaller than that of the case $\sigma=+1$
for a fixed $j$ value. Hence a larger value of $j$ is needed for
the disappearance of the AH. 
This explains our result $j_{\rm crit}^{(-)}(L) \ge  j_{\rm crit}^{(+)}(L)$.

%
%
\section{Summary and discussion}

In this paper, we studied the AH formation in the head-on collision of
gyratons. 
We introduced four gyraton models in Sec. II:
a spinless $p$--gyraton, an AS-gyraton, and spinning {\it a}--
and {\it b}-- gyratons.
The energy and spin profiles of each gyraton are given 
in Eqs.~\eqref{chip-chij-withoutspin}--\eqref{chip-chij-typeb}.
For a spinless $p$--gyraton and an AS-gyraton, the energy profile 
is a step function with width $L$ and a delta function, respectively. 
For {\it a}-- and {\it b}-- gyratons,
the energy profile is a delta function and the spin profile is a step function
with width $L$.
The difference between {\it a}-- and {\it b}-- gyratons is the 
relative locations of the energy and spin profiles. 
We introduced the null geodesic coordinates for each
gyraton, and discussed the property of its gravitational field. 
Especially, a spinning gyraton has a repulsive gravitational field
around its center.

%
\begin{table}[tb]
\centering
\caption{Summary of the obtained results. 
For each case, the condition of AH formation was found
in terms of $L$ [the energy duration in the (0) case
and the spin duration in other cases] and the spin value $j=J/2pr_h(2p)$ 
(assumed to be positive). The unit of the length is $r_h(2p)=4Gp$.}
\begin{ruledtabular}
\begin{tabular}{ccccll}
collision type & slice ($\sigma$) &   gyraton 1 & gyraton 2 
& \multicolumn{2}{l}{condition of AH formation}  \\ \hline
(0) & $\cdots$ & $p$ & $p$ & \multicolumn{2}{c}{$L\lesssim 1.4$}~~ \\
(1{\it a}) & $\cdots$ & {\it a} & AS & ~~$L\sim 1$ & ~~$j\lesssim 0.25$ \\
(1{\it b}) & $\cdots$ & {\it b} & AS & ~~$L\sim 1$ & ~~$j\lesssim 0.25$ \\
(2{\it a}) & $\cdots$ & {\it a} & {\it a} & ~~$L\sim 1$ & ~~$j\lesssim 0.2$ \\
(2{\it b}) & Old & {\it b} & {\it b} & ~~$L\sim 0.15$ & ~~$j\lesssim 0.1$ \\
(2{\it b}) & New ($+1$) & {\it b} & {\it b} & ~~$L\sim 0.5$ & ~~$j\lesssim 0.1$ \\
(2{\it b}) & New ($-1$) & {\it b} & {\it b} & ~~$L\sim 0.5$ & ~~$j\lesssim 0.15$ \\
  \end{tabular}
  \end{ruledtabular}
  \label{summary}
\end{table}
%

Then the problem of the head-on collisions of two gyratons 
was set up and
the AH was studied on the slice $u=0\ge v$ and $v=0\ge u$ 
in Secs. II--IV.
The studied collision cases and obtained results are summarized
in Table~\ref{summary}. 
In all cases two gyratons are assumed to have the same energy.
Case (0) is the collision of two identical
spinless $p$--gyratons. In this case the energy duration $L$
should be smaller than some critical value for the AH formation. 
In cases (1{\it a}) and (1{\it b}), we studied the collision of 
spinning {\it a}-- and {\it b}-- gyratons with an AS-gyraton, respectively.
We obtained the conditions for the AH formation  
in terms of the spin duration $L$ and the spin $j$. 
They are shown in Fig.~\ref{region-case1}
and roughly summarized as in Table~\ref{summary}.
(Here $j>0$ is assumed 
since the AH formation does not depend on the spin direction.) 
In both cases, there was a critical value $j_{\rm crit}(L)$ for the AH formation
for a given $L$. We found no significant difference between the two cases.
In cases (2{\it a}) and (2{\it b}), we studied the collision of two 
spinning {\it a}-- and {\it b}-- gyratons, respectively.
Two gyratons were assumed to have the same spin duration $L$
and absolute value of the spin $j$. 
We obtained the conditions for AH formation 
in terms of $L$ and $j$.
They are shown in Fig.~\ref{region-case2} and 
roughly summarized as in Table~\ref{summary}.
(Here $j>0$ is assumed and the relative direction of two spins $\sigma$
is not specified, since the AH formation does not depend on the directions 
of two spins on the studied slice.)
We found that the allowed region on the $(L,j)$-plane in the (2{\it b}) case
is much smaller than that in the (2{\it a}) case.

In Sec. V, we focused our attention on
the gravitational spin-spin interaction after collision in the (2{\it b}) case.
We solved a part of the future to the slice $u=0\le v$ and $v=0\le u$
(old slice) in the collision of gyratons with spins $j$ and $\sigma j$ ($\sigma=\pm 1$),
using a method of perturbation where $j$ is a small expansion parameter.
The solved region is the past to the
collision of the energy flows, but the two spin flows
interact with each other in that region (see Fig.~\ref{region-IVA} for details).
Therefore we could study the spin-spin interaction.
Then we again studied the AH formation on 
the future edge of the solved region (new slice) and 
compared the obtained results to those of the old slice.
It was found that the allowed region becomes
larger by just changing the slice (Fig.~\ref{secondorder-jmj}).  
Hence, the difference between the results of 
old slice in cases (2{\it a}) and (2{\it b}) was due to
the artificial effect of choosing a slice on which we study the AH.

Furthermore, we found the dependence on the relative helicities of
incoming gyratons. In the case $\sigma=+1$ where
two spins have the same direction (i.e., helicities have opposite signs),
the gravitational field is repulsive everywhere due to the spin-spin interaction.
On the other hand, in the case $\sigma=-1$ where two spins 
have the opposite directions
(i.e., helicities have the same sign), the spin-spin interaction decreases the repulsive force
and even changes it into the attractive force in a part of the studied region. 
Correspondingly the allowed region of the AH formation for
$\sigma=-1$ is larger than that for $\sigma=+1$ 
(Fig.~\ref{secondorder-jmj}).

In the light of the above studies, we claim the following. For the
AH formation in the head-on collision of gyratons,
(i) the energy duration should be smaller than some critical value 
(close to the system gravitational radius);
(ii) the spin duration should be at least 
of order of the system gravitational radius
(it should not be too small or too large);
(iii) the spin should be smaller than some critical value 
that is a function of the spin duration.
Further, 
(iv) the AH formation in the collision of two gyratons 
with the oppositely directed spins
is easier than that with the same direction of spins.

Now we discuss the possible applications of the obtained results
for mini black hole production at the LHC in the TeV gravity scenarios
where $M_p=\mathrm{TeV}$.
Let us consider the collision of two spinning particles,
and use our result of the (2{\it a}) case, i.e. the 
collision of two identical {\it a}--gyratons, 
for the condition for the black hole formation as an example.
Restoring the length unit, it is written as $L\simeq r_h(2p)$
and $J\lesssim 0.4\times p \ r_h(2p)$.
We use the Lorentz contracted proton size 
$L \sim 1.5\times 10^{-4}\mathrm{fm}$ for the spin duration
and put $J=\hbar/2$ as possible candidates for these values. 
Substituting
$p=\mathrm{(few)}M_p$ and $r_h(2p)=\mathrm{(few)}\hbar/M_p$,
we find $L\sim r_h(2p)$ and
%
\begin{equation} 
0.4\times p \ r_h(2p)\sim\mathrm{(few)}\hbar
\gtrsim \hbar/2=J.
\end{equation}
%
Hence, the above two conditions are satisfied
and the black hole is expected to form in the head-on collision
under our assumption.
Thus, the effect of spins of incoming particles might not be significant
for the black hole production rate. Still, the spin might change
the cross section of the black hole production by a factor
and studying this effect would be interesting.

We also revisit the study by
Giddings and Rychkov \cite{GR04}, because
our study is related to the assumption they made.
In that paper the collision of quantum wave packets with width $L$  was considered.
Their result is that if $\hbar^2/(r_hM_p^2)\ll L\ll r_h$, 
the higher-curvature correction 
is small and the predictions by general relativity
are reliable.
The latter inequality $L\ll r_h$ was imposed by
the expectation that the gravitational field of such a wave packet
would be sufficiently close to that of the AS particle and thus 
the AH would form in a collision of such two wave packets.
Our result of the (0) case, i.e. the collision of two identical $p$--gyratons,
explicitly demonstrates the accuracy of this expectation.
Moreover, because we found the AH also for $L\lesssim 1.4r_h$,
the condition can be relaxed
to $\hbar^2/(r_hM_p^2)\ll L\lesssim r_h$.
Note that this criterion holds also for wave packets of spinning particles,
if their energy is sufficiently large, $p\gg M_p$.
Our results of the (2{\it a}) and (2{\it b}) cases, i.e. the collisions of two
identical spinning {\it a}-- and {\it b}-- gyratons, show that
$j^2\lesssim L$
is necessary for the AH formation for small $L$.
Restoring the length unit and adopting $J=\hbar/2$, 
it is rewritten as $\hbar^2/(16r_hp^2)\lesssim L$.  
However this does not provide a new condition
since the original condition $\hbar^2/(r_hM_p^2)\ll L$
implies $\hbar^2/(16r_hp^2)\lesssim L$ for $p\gg M_p$.
Therefore, our results do not contradict the claims in \cite{GR04}.

The important remaining problems are as follows.
The first one is to explore the case $\sigma=-1$ further.
This is because the condition of the black hole formation 
is expected to be different from that of the AH formation. 
In the case $\sigma=+1$, however, 
the critical value of $j$ for the black hole formation
will remain finite, because both the gravitational field generated
by the spin source and the spin-spin interaction
are repulsive. On the other hand,
in the case $\sigma=-1$, 
the repulsive gravitational field 
of each incoming gyraton is weakened and becomes
even attractive in some part of the spacetime
by the spin-spin interaction
as shown in Sec. IV. Hence, there is 
the possibility that later the gravitational field turns to be attractive everywhere
and the critical value of $j$ blows up.

The next problem is the collision of gyratons with a nonzero impact parameter.
In these grazing collisions, 
new effects of the spin-orbit interaction will appear. 
Moreover, the properties of spin-spin interaction might change.
Let us recall Eq.~\eqref{spin-spin-force}, the acceleration $\vec{a}_{\rm SS}$
due to the spin-spin interaction between weakly gravitating bodies. 
In the grazing collisions, the spins are orthogonal to the relative location vector 
and $\vec{a}_{\rm SS}$ is calculated as
$\vec{a}_{\rm SS}=-(3/\mu r^4)(\vec{S}_1\cdot\vec{S}_2)\vec{n}$.
Therefore in the aligned (resp. anti-aligned) case,
the spin-spin interaction becomes attractive (resp. repulsive), which 
is opposite to the head-on collision case. 
Therefore it is expected that the nonzero impact parameter
would make the interactions more complicated but more interesting.

It is also important to simulate the collision of gyratons with
realistic sources. In this paper, we assumed that each incoming
gyraton has a singular source, Eqs.~\eqref{Tuu} and \eqref{Tua},
and studied only the spacetime regime where 
the matter interaction is not important (i.e., $uv<r^2$ in Sec. V).
In a realistic situation, however, the source of an incoming gyraton
is a beam pulse with a finite radius $\bar{r}_{\rm s}$. 
Then, the matter interaction determines the spacetime structure
within the lightcone of the source collision, and the condition
for the black hole formation will depend on the 
properties of matter interaction. In order to study
this effect, we should solve the Einstein equations together
with the field equations for the sources.

Finally, the generalization for the higher-dimensional case
is necessary to obtain the results that can be directly applied
for the black hole production at accelerators in the TeV gravity scenarios.

%
%
\acknowledgments

H.Y. thanks Tetsuya Shiromizu for helpful disscussions
at the early stage of this work.
The authors thank the
Killam Trust for financial support.  
One of the authors (V.F.) is greatful to the NSERC 
for its support.

\end{document}